\documentclass{article}

\usepackage[numbers,sort&compress]{natbib}

\usepackage{amssymb}
\usepackage{amsmath}
\usepackage{graphicx}
\usepackage{amsthm}
\usepackage{bm}
\usepackage{enumitem}
\setlist[enumerate]{label=(\arabic*)}

\newcommand{\vctr}[1]{\bm{ #1 }}

\newcommand{\dr}[1]{\ensuremath{\mathrm{d} #1\,}}
\newcommand{\ddt}{\ensuremath{\frac{\dr{}}{\dr{t}}}}
\newcommand{\dbd}[2]{\ensuremath{\frac{\dr{#1}}{\dr{#2}}}}
\newcommand{\dbdbd}[2]{\ensuremath{\frac{\mathrm{d}^2{#1}}{\dr{#2}^2}}}
\newcommand{\pbp}[2]{\ensuremath{\frac{\partial #1}{\partial #2}}}
\newcommand{\pbpbp}[3]{\ensuremath{\frac{\partial^2 #1}{\partial #2 \partial #3}}}

\newcommand{\tr}{\textup{\textsf{Tr}}}

\newcommand{\be}{\begin{equation}}
\newcommand{\ee}{\end{equation}}

\newtheorem{theorem}{Theorem}
\newtheorem{lemma}{Lemma}
\numberwithin{equation}{section}
\begin{document}

\title{A Gravitomagnetic Bootstrap: from Newtonian gravity to general relativity via localizing rotational and conformal symmetries}
\author{David Wallace\thanks{Department of History and Philosophy of Science / Department of Philosophy, University of Pittsburgh; \texttt{david.wallace@pitt.edu}}}
\maketitle

\begin{abstract}
Newtonian gravity, in the light of the Equivalence Principle, is best understood as a localization of Galilean inertial structure --- or at least, of the part of Galilean inertial structure that defines rectilinear acceleration. I consider what kind of theory can be constructed if we also localize rotational inertia and spatial and temporal scales. The resultant theory, given weak assumptions, is unique up to a constant with the dimensions of velocity and an overall sign. For an appropriate choice of these, it coincides with the first post-Newtonian approximation to general relativity, so that this construction gives a method to bootstrap from Newtonian gravity to general relativity without explicitly presupposing special relativity.
\end{abstract}
\section{Introduction}\label{introduction}

In popular accounts, general relativity is often presented as a radical conceptual break from Newtonian gravity, replacing a theory of long-distance forces acting within a background of absolute space and time with a theory of dynamical spacetime geometry. But the central insight that led Einstein to general relativity --- the Equivalence Principle, the equivalence between physics in an accelerating frame and physics under the action of a constant gravitational field --- was not something Einstein learned through intuition or observation: it was an insight about Newtonian gravity itself.\footnote{For further development of this observation, including the important point that we do not need to change the formalism of Newtonian gravity in order to interpret it geometrically in this way, see, e.g., \citep{knoxequivalence,wallacenewton}.}

Einstein realized that Newtonian gravity could be better understood not as a theory of long-distance forces but as a theory that localized inertial structure. And since he had already concluded that Newton's picture of inertial structure was not quite right and had to be replaced with the version implicit in \emph{special} relativity, the path to the general theory was ``just'' a matter of working out how to marry those two insights.

But it seems to be a matter of historical contingency that this insight into the nature of Newtonian gravity was gained only after physicists had already learned the limitations of Newton's and Galileo's conception of inertia and spacetime. (Indeed, Newton seemed at least to glimpse it: as argued by, e.g.,  Saunders~\citep{SaundersPrincipia}, Corollary VI to the \emph{Principia} is a clear precursor to the Equivalence Principle.) And had physicists come to it sooner, they might also have noticed that the localization of inertial structure implicit in Newtonian gravity is only partial. The standard of acceleration might be localized and made dependent upon the matter distribution, but the standards of rotation, and of spatial and temporal scales, remain global and absolute. And they might then have wondered what kind of theory could result if those aspects of inertia were also made dynamical.

This paper is an exploration of that possibility (an exploration, to be clear, that I intend as an anachronistic thought-experiment and not as a serious piece of counterfactual history). As I will show, any such theory is extremely tightly constrained, indeed so tightly constrained that it is almost uniquely fixed up to a new constant of nature with the dimensions of velocity. And to make the resultant theory self-consistent it is necessary to modify the Galilean symmetries of Newtonian physics in exactly the way that proved necessary in special relativity: that is, to replace them with the Lorentz symmetries, with the new constant of nature playing the role of the speed of light. The resultant extension of gravity can then be understood as the first post-Newtonian approximation to general relativity, and the conclusion is that we can bootstrap from the initial hypothesis of localized rotation and scale up to general relativity, in a way strikingly similar to the well-known bootstrap from magnetostatics to the Lorentz transformations and the Maxwell equations. 

After reviewing the magnetostatic bootstrap (section~\ref{bootstrap}), I begin (section~\ref{localizedinertia}) by considering what kind of theory can be made by localizing the Newtonian standard of unaccelerated motion, and argue that under weak assumptions the form of that theory is fixed uniquely up to the magnitude and sign of the gravitational constant. I then consider what kind of theories can be constructed under those same weak assumptions from localizing more general aspects of inertial structure. My first attempt (section~\ref{generalizinginertial}) localizes both translational (that is: rectilinear-acceleration) and rotational structure, introducing a `Coriolis field' to represent the latter; unfortunately, the resultant theory is inconsistent unless the Coriolis field is source-free. The second attempt (section~\ref{localizedscale}) also localizes spatial scale structure; this turns out to fix the inconsistency in the previous theory and to produce a very rich extension of Newtonian gravity, uniquely defined up to the gravitational constant and a new constant with the dimensions of velocity. This extension, however, fails to be Galilean-covariant, but if we also localize the temporal scale (section~\ref{temporalscale}) it can be made covariant under a new symmetry which agrees with Galilean boost symmetries in the limit of low velocities. The resultant theory (section~\ref{1pntheory}) is the first post-Newtonian approximation to general relativity (section~\ref{bootstrapping-gr}). I conclude (section~\ref{consequences}) by exploring some conceptual, philosophical, and pedagogical lessons of this construction. 

Notation: Roman indices range over $(1,2,3)$ and are raised and lowered freely; repeated indices are summed over; if $\vctr{v}$ is a vector, $v^2\equiv \vctr{v}\cdot\vctr{v}$; if $\vctr{T}$ is a tensor and $\vctr{v}$ a vector, $(\vctr{T}\cdot\vctr{v})_i\equiv T_{ij}v_j$. For readability, most of the purely mathematical results used are proved in the Appendix, which is largely self-contained.

\section{Recap: the magnetostatic bootstrap}\label{bootstrap}

(This is all well-known textbook material and I claim no originality.)

Start with the equations for \emph{(electro)-magnetostatics}, the (classically exact) theory of how test charges move in the presence of time-independent charges and currents. These are the Lorentz force law,
\be
\dbd{(m\vctr{v})}{t}=q (\vctr{E}+\vctr{v}\times \vctr{B}),
\ee
and the static Maxwell equations,
\begin{subequations}
\begin{align}
\vctr{\nabla}\cdot{\vctr{E}}&=\frac{\rho}{\epsilon_0} \label{ms-eq-1}\\
\vctr{\nabla}\cdot{\vctr{B}}&=0\label{ms-eq-2}\\
\vctr{\nabla}\times\vctr{E}&=\vctr{0} \label{ms-eq-3}\\
\vctr{\nabla}\times\vctr{B}&=\mu_0 \vctr{J}.\label{ms-eq-4}
\end{align}
\end{subequations}
Under these equations, given two infinitesimal pieces of moving charge with velocities $\vctr{v}_1$, $\vctr{v}_2$ and charges $q_1$, $q_2$, the electric force between them scales like $(q_1 q_2/r^2)(1/\epsilon_0)$ and the magnetic force scales like $(q_1 q_2/r^2)(\mu_0 |\vctr{v}_1||\vctr{v}_2|)$, so that $\mu_0\epsilon_0$ must have the dimensions of inverse velocity squared, $\mu_0\epsilon_0=1/c^2$ for some velocity $c$.

Taking the divergence of \eqref{ms-eq-4} gives $0=\mu_0\vctr{\nabla} \cdot \vctr{J}$. Local conservation of charge is expressed by $\partial_t \rho + \vctr{\nabla} \cdot \vctr{J}=0$, so that assuming charge is conserved, $\partial_t\rho=0$. Allowing time-varying charge density thus requires adding some time-dependent term to the right-hand side
of \eqref{ms-eq-4} whose divergence is $\mu_0\partial_t\rho$, and it is easy to see from \eqref{ms-eq-1} that this modification must be 
\be\label{ms-eq-4-m}
\vctr{\nabla}\times\vctr{B}=\mu_0 \vctr{J}+\mu_0 \epsilon_0 \partial_t\vctr{E}.
\ee
Our theory may or may not be Galilean-invariant, but at any rate failure of Galilean invariance had better appear only at order $1/c^2$ or higher, otherwise experiments in labs on different sides of the planet would give detectably different results. If for the moment we assume exact  invariance of the Lorentz force law under $\vctr{x}\rightarrow \vctr{x}+\vctr{u}t$, this requires the transformation law
\begin{subequations}
\begin{align}
\vctr{E}&\rightarrow \vctr{E}-\vctr{u}\times \vctr{B} \\
\vctr{B}&\rightarrow \vctr{B}.
\end{align}
\end{subequations}
Under this transformation we have
\begin{align}
\vctr{\nabla}\times \vctr{E}&\rightarrow \vctr{\nabla}\times \vctr{E}-\vctr{\nabla}\times (\vctr{u}\times \vctr{B})\nonumber\\
&=\vctr{\nabla}\times \vctr{E}-\vctr{u}\vctr{\nabla}\cdot\vctr{B}+ (\vctr{u}\cdot\vctr{\nabla})\vctr{B}
\end{align}
so that \eqref{ms-eq-3} is not Galilean-invariant even if \eqref{ms-eq-2} is assumed. Fixing this requires replacing \eqref{ms-eq-3} with
\be
\vctr{\nabla} \times \vctr{E}+\partial_t \vctr{B}=0
\ee
which is the last Maxwell equation.

However, \eqref{ms-eq-1} and \eqref{ms-eq-4-m} are invariant under Galilean transformations only up to terms of order $1/c^2$. To preserve exact invariance of the Maxwell equations under boosts we have to replace the Galilei group with the Lorentz group and change the transformation law for magnetic fields to
\begin{align}
\vctr{B}&\rightarrow \vctr{B} + \mu_0 \epsilon_0 \vctr{u}\times\vctr{E}
\end{align}
(note that again this is the Galilean form up to a term of order $1/c^2$).  The Lorentz force law is in turn invariant under these transformations only if we assume mass depends on velocity like $m(v)=m_0 (1-v^2/c^2)^{-1/2}$. Now we are done: we have proceeded from magnetostatics and the assumption of leading-order Galilean symmetry, to the full Maxwell equations, the Lorentzian symmetries of special relativity, and the velocity-dependence of inertial mass.

\section{Newtonian gravity as localized inertia}\label{localizedinertia}

Now  let us forget what we know of electrodynamics and relativity, and return to Newtonian gravity.

\subsection{Newtonian dynamics and global inertial structure}

Start with some collection of $N$ point particles with masses $m_1,\ldots,m_N$ moving in Euclidean space, interacting by position-dependent forces: the force exerted by particle $n$ on particle $k$ is $\vctr{F}_{kn}$, acts along the line between $n$ and $k$, and has magnitude dependent only on the distance $|\vctr{x}_n-\vctr{x}_k|$ between the spatial locations of the particles. By Newton's Third Law, $\vctr{F}_{kn}=-\vctr{F}_{nk}$, and the equations of motion for the system are
\be
\dbdbd{\vctr{x}_n(t)}{t} = \frac{1}{m_n}\sum_{k\neq n} \vctr{F}_{nk}(\vctr{x}_n(t)-\vctr{x}_k(t)).
\ee
This system of equations is Galilean-covariant: solutions are preserved under the transformations
\begin{subequations}
\begin{align}
\vctr{x}(t)&\rightarrow \vctr{R}\vctr{x}(t) + \vctr{a} + \vctr{u} t;\\
t&\rightarrow t + \tau
\end{align}
\end{subequations}
for any time-independent rotation matrix $\vctr{R}$, vectors $\vctr{a},\vctr{u}$, and scalar $\tau$. They are not, however, invariant under the wider class of time-dependent translations $\vctr{x}\rightarrow \vctr{x}+\vctr{a}(t)$ (for an arbitrary smooth vector-valued function $\vctr{a}(t))$, nor the still-wider class of time-dependent rotations-plus-translations, where $\vctr{R}$ is allowed to depend on time. Insofar as we take the symmetries of the equations as characterizing the structure of spacetime, this is to say that the implied setting of this dynamical system is Galilean (or neo-Newtonian) spacetime, not Maxwellian spacetime (which allows arbitrary time-dependent translations) or Leibnizian spacetime (which also allows time-dependent rotations).\footnote{My nomenclature follows~\citep{earman}. The Maxwell symmetry group is also called the \emph{Milne group}~\citep{CarterKhalatnikov1994}; the Leibniz group is sometimes called the \emph{Coriolis group}~\citep{Duval1993}.}

Another way to characterize the covariance of these equations is that they hold in arbitrary \emph{inertial frames}, where the inertial frames are those in which force-free bodies move in straight lines at constant speed; given one inertial frame, performing a Galilean transformation yields another such frame. As written, they do \emph{not} hold in accelerated frames --- but formally we can fix this easily enough, just by introducing a new vector field  $\vctr{G}(t)$ and rewriting the equations as
\be\label{acceleratedmotion}
\dbdbd{\vctr{x}_n(t)}{t} - \vctr{G}(t)= \frac{1}{m_n}\sum_{k\neq n} \vctr{F}_{nk}(\vctr{x}_n(t)-x_k(t)).
\ee
The new system of equations is invariant under the wider \emph{Maxwell} symmetry group 
\begin{subequations}\label{maxwell-group}
\begin{align}
\vctr{x} (t)&\rightarrow \vctr{R}\vctr{x} (t)+ \vctr{a} (t)\\
t&\rightarrow t + \tau\\
\vctr{G}(t)&\rightarrow \vctr{R}\vctr{G}(t)+\dbdbd{\vctr{a}(t)}{t}.
\end{align}
\end{subequations}
A useful way to think about this is that a Galilean-covariant dynamical system has an implied \emph{inertial structure}, which suffices to distinguish unaccelerated and unrotated motion from accelerations and rotations. In introducing $\vctr{G}$, we have represented the accelerational part of the inertial structure by an \emph{inertial force}: the inertial force on each body is equal to its mass times $\vctr{G}$, it can be treated just as the `real' forces $\vctr{F}_{nk}$, and the inertial frames are just those in which $\vctr{G}$ vanishes.

\subsection{Localizing translational structure}

In itself, this move tells us little substantive about the dynamical system: the move is simply a special case of Kretschmann's famous observation~\citep{Kretschmann1917,nortoneightdecades} that any theory can be put in a generally covariant form with enough mathematical care. And the symmetry here is purely formal, and does not correspond to any \emph{physical} symmetry (that is, to any symmetry that can be performed on an isolated subsystem of a larger system, as in the case of Galileo's ship\footnote{Here I presume a general framework for discussing the physical content of symmetries that I develop in more detail in~\citep{wallaceobservingsymmetries}: a symmetry is physical if it can be regarded as a transformation of some subsystem, describable against some background which does not transform.}).

However, the same mathematics can be reinterpreted in a way which does tell us something substantive. To begin, let's define a \emph{force field} as a field on a region of spacetime which induces some force on each body in the region, where that force might depend on the body's location and velocity and also on intrinsic properties like charge or mass. And a force field is \emph{universal} if the force it induces on a (localized) body brings about an acceleration that depends perhaps on the body's location and/or velocity but at any rate not on any other feature of the body.

One such force field is specified by our vector function $\vctr{G}(t)$: if we declare that it induces a force 
\be
\vctr{F}_n[\vctr{G},t]=m_n \vctr{G}(t)
\ee on the $n$th particle, then the combined effect of that force on the system of particles (together with the interparticle forces) is to induce exactly the equations of motion \eqref{acceleratedmotion}. Which is to say: a spatially invariant, velocity-independent \emph{universal} force is indistinguishable from the inertial force induced by a time-dependent translation into an accelerating reference frame; put another way, universal forces of this form can be reinterpreted as part of the inertial structure. This is a form of Einstein's Equivalence Principle (see \citep{lehmkuhlequivalence}) or of Newton's Corollary VI (see \citep{SaundersPrincipia}).

Now suppose that we allow $\vctr{G}$ to vary with space as well as time --- but that we require its spatial variation to be slow enough in the vicinity of the particles that the relative accelerations between particles $n$ and $p$ induced by the universal force are negligible compared to those induced by interparticle forces: that is,
\be
\vctr{G}(\vctr{x}_n)-\vctr{G}(\vctr{x}_p) \ll \frac{1}{m_n}\left(\sum_{k\neq n}\vctr{F}_{nk}\right) - \frac{1}{m_p}\left(\sum_{k\neq p}\vctr{F}_{pk}\right)
\ee
(the exact conditions for this will depend on the configuration of the particles, on the form of the interparticle forces, and of course on our tolerance for approximation). Under these conditions, the particles move under their interparticle accelerations together with a shared universal acceleration $\vctr{G}(\vctr{X})$, where $\vctr{X}$ is some point characterizing the overall location of the particle swarm (say, the center of mass). We now have a stronger result (a form of the \emph{Strong Equivalence Principle}; see again \citep{lehmkuhlequivalence}):
\begin{quote}
Any universal and velocity-independent force field is indistinguishable \emph{locally} from a shift of inertial structure or a shift to a rectilinearly accelerating reference frame, provided it is probed on scales where the interparticle accelerations of the  probing matter due to internal forces are large compared to the relative acceleration of different particles induced from their differing locations and the spatial variation of the universal force field.
\end{quote}
So a theory characterized by a universal velocity-independent force field $\vctr{G}$ is indistinguishable from a theory in which inertial structure --- or at least that part of inertial structure which distinguishes uniform rectilinear acceleration from inertial motion --- is localized and can vary from place to place. And by `indistinguishable' I mean not just that no \emph{experiment} can distinguish the possibilities, but that they are identical at the level of mathematical formalism. The same vector field $\vctr{G}$ can be interpreted either as a universal force on Galilean spacetime, or as a local standard of what motions count as force-free.

I am inclined to say (see \citep{wallacemathfirst}) that the question of which interpretation is `correct' is a pragmatic rather than factive matter; following Knox~\citep{knoxequivalence} I am inclined to add that the pragmatically sensible choice is to take spacetime structure as by definition given by the motions of bodies under inertial forces, so that the `correct' spacetime structure is the one in which inertial motion is localized, and not the unobservable background Galilean spacetime. But my concern here is less whether $\vctr{G}$ has a univocal interpretation, more what kind of dynamical theory we can build that incorporates such a universal force field.

\subsection{Deriving Newtonian gravity}

Attempts to lock down the form of a physics theory are usually heuristic rather than deductive: one states reasonable-sounding assumptions and sees what follows from them, without a claim that the assumptions are logically necessary. With this modest understanding of the game we are playing, let us make the following assumptions about our theory.
\begin{enumerate}
\item The force field $\vctr{G}$ is determined by the matter distribution (that is, by where the matter is, how fast it is going, and perhaps other intrinsic properties like mass, charge or shape). This does not entirely preclude the possibility that $\vctr{G}$ has independent degrees of freedom (analogous to electromagnetic radiation) but it does presume that we can identify a regime where those additional degrees of freedom are absent. 
\item The dynamics for $\vctr{G}$ are local at least inasmuch as they are given by some local differential equations that relate $\vctr{G}$ and its spatial and temporal derivatives to some local function of the matter distribution. Furthermore, as a simplifying \emph{ansatz} rather than for any deeper reason, we will assume these equations are \emph{first order} and that the matter distribution occurs linearly in them. Given the additional assumption of Euclidean symmetry which we make below, this amounts to saying that the equations constrain the divergence, curl, and rate of change of $\vctr{G}$, equating them to matter densities and currents (or to zero, in the absence of matter).
\item The equations in (2) are solved for empty space in accelerated coordinates: that is, for uniform $\bm{G}$.
\item Given some collection of particles very distant from other particles, we can treat that collection as an isolated system and solve for the $\vctr{G}$ associated to that system, under the assumption that it tends to zero at large distances. This is basically a scale separation assumption: there may in addition be a `background' $\vctr{G}$-field attributable to distant matter, but provided that it varies slowly enough that it can be treated as constant over the extent of the system, it can be transformed away by boosting to an accelerating reference frame.
\item The equations of the theory are covariant under the Maxwell group, with the transformation law for the field being \eqref{maxwell-group}, the one appropriate for an inertial field.
\item The overall dynamics of matter under gravity conserves momentum (if we assumed an action formalism we could regard this as a consequence of (5), given that momentum is the Noether invariant associated to spatial translation symmetry, but we will not do so).
\end{enumerate}
Assumptions (1)--(6) pretty much pin down the form of the theory. To see this, firstly note that uniform fields have vanishing curl and divergence. By (1)--(3), that fixes the field equations in vacuum to be
\begin{subequations}
\begin{align}
\vctr{\nabla} \cdot \vctr{G}&=0\\
\vctr{\nabla} \times \vctr{G}&=0
\end{align}
\end{subequations}
Now, suppose we have some region containing matter, with mass density $\rho$. The force per unit volume on some small chunk of the matter is 
\be\label{force-per-unit}
\vctr{f}=\rho \vctr{G}.
\ee
That is: the force on a small chunk of matter depends only on its mass and not on its velocity or anything specific to the type of matter the chunk is made up of (such as electric charge or chemical composition). If we then consider (say) two rocks interacting under gravity, then the gravitational force \emph{on} each rock depends only on that rock's mass, and so in order to conserve the total momentum of the system, the gravitational force exerted \emph{by} each rock can depend only on its mass. This entails that the gravitational force must be sourced by $\rho$ alone, and given our other assumptions, the only available  field equations are
\begin{subequations}\label{newton-grav-continuum}
\begin{align}
\vctr{\nabla} \cdot \vctr{G}&=-4\pi G_N \rho\\
\vctr{\nabla} \times \vctr{G}&=0
\end{align}
\end{subequations}
We can readily verify that the theory is indeed Maxwell-covariant. The constant $G_N$ is a free parameter, and indeed nothing about our arguments requires it to be positive: a localized theory of acceleration must be universally repulsive or universally attractive, but empirical data is required to fix which.

If we consider some distribution of matter such that the background gravitational field can be idealized as constant over the extent of the matter, and we then perform a coordinate transformation to eliminate that background field, we can adopt boundary conditions where $\vctr{G}$ tends to zero at spatial infinity; with this boundary condition, $\vctr{G}$ is given explicitly by the familiar Newtonian expression
\be\label{phi-explicit}
\vctr{G}=\vctr{\nabla} \Phi;\,\,\,\,\,\Phi(\vctr{x})=G_N \int\dr{^3y}\frac{\rho(\vctr{y})}{|\vctr{x}-\vctr{y}|}.
\ee
For a point source, of course, this reduces to the inverse square law: the gravitational force due to a particle of mass $m$ has magnitude $G_N m/r^2$ at a distance $r$ from that particle.

To conclude: on fairly weak and plausible additional assumptions, and up to the sign and magnitude of Newton's constant, the form of Newtonian gravity is pretty much determined once we require it to be the theory of a universal velocity-independent force --- or equivalently, once we require it to be a theory that localizes the part of global inertial structure that distinguishes uniform acceleration from inertial motion. And the overall system has as its symmetries the Maxwell group of Euclidean transformations, time translations, and time-dependent spatial translations.

\subsection{Continua, extended bodies, and the Equivalence Principle}\label{continuumsection}

My account so far has moved back and forth between characterizing matter as discrete particles and as a continuum, and it will be helpful for later to pause and say a little about how these are related.
In astrophysics, `particle' does not refer to genuinely pointlike singularities, nor `continuum' to matter genuinely continuous at all scales. Rather: these are intrinsically approximate and scale-dependent notions. In both cases, the important notion is scale separation: a body of matter can be treated as a point particle if the scale on which it is extended is very small compared to the other dynamically relevant scales in the problem (such as: how quickly the background fields vary). Similarly, a collection of many such particles can be treated as a continuum if we can identify `representative volume elements' (RVEs), volumes large enough to contain a great many particles but still small compared to the other relevant scales of the problem.

(This scale relativity of matter is a commonplace of astrophysics. On kiloparsec scales, the Galaxy can be treated as a fluid of stars. Zoom in, and that fluid resolves into individual stars which can be treated as point particles. Zoom in further, and those point particles resolve into a fluid of plasma. Zoom in further still, and that fluid resolves into electrons and nuclei, albeit at this scale classical mechanics itself breaks down.)

Thus understood, the continuum dynamical equations can be understood as consequences of the discrete ones. In point-particle mechanics, the basic dynamical principles are that force is rate of change of momentum and that individual particle mass is conserved. The continuum versions of these are
\begin{subequations}
\label{eq:contgeneral}
\begin{align}
\label{eq:contgeneral-force}
\partial_t \vctr{J}+\vctr{\nabla} \cdot \vctr{T}&=\vctr{f}\\
\partial_t \rho+\vctr{\nabla} \cdot \vctr{J}&=0 \label{eq:contgeneral-mass} 
\end{align}
\end{subequations}
where $\rho$, $\vctr{J}$, $\vctr{T}$, and $\vctr{f}$ are respectively the mass density, momentum density, stress tensor, and external force density. If we understand the continuum as a swarm of particles then these are defined by
\begin{subequations}
\begin{align}
\rho \times \mbox{volume of RVE}&= \sum_n m_n  \\
J_i \times \mbox{volume of RVE}&=\sum_n m_n v^n_i  \\
T_{ij}\times \mbox{volume of RVE}&=\sum_n \left ( x^n_j F^{n,\mathrm{int}}_i + m_n v^n_i v^n_j\right)\\
f_i \times \mbox{volume of RVE}&=\sum_n F^{n,\mathrm{ext}}_i
\end{align}
\end{subequations}
where the sum in each case is over the particles in the RVE and $m_n,\vctr{x}^n$, $\vctr{v}^n$, $\vctr{F}^{n,\mathrm{int}}$, $\vctr{F}^{n,\mathrm{ext}}$ are respectively the mass, position, and velocity of the $n$th particle in the RVE and the forces on that particle from other particles in the RVE and from outside the RVE. As a special case, in a gravitational field $\vctr{G}$ we have
\be
\vctr{f} \times \mbox{volume of RVE}=\sum_n \vctr{F}^{n,\mathrm{ext}}=\vctr{G}\sum_n m_n=\vctr{G}\rho  \times \mbox{volume of RVE}
\ee
so that $\vctr{f}=\rho\vctr{G}$, in agreement with \eqref{force-per-unit}.

However, we can equally go the other way around. If we take our theory to be defined by the continuum equations~\eqref{force-per-unit}, \eqref{newton-grav-continuum}, \eqref{eq:contgeneral-mass}--\eqref{eq:contgeneral-force}, we can consider some extended (but finite) chunk of continuous matter held together by internal forces, which could be non-gravitational (e.g., a star) or gravitational (e.g., a star cluster). Integrating \eqref{eq:contgeneral-force} over the extent of the body recovers Newton's Second Law: 
\be
\int \vctr{f}=\ddt \int \vctr{J}.
\ee
The mass of the body can be defined in several logically distinct ways:
\begin{itemize}
\item The \textbf{active (gravitational) mass} $M^A$ is defined by
\be
-4\pi G_N M^A=\oint \vctr{G}\cdot \mathrm{d}\vctr{S}
\ee
where the integral is taken over a surface surrounding the body (in the sequel we will require it to be very far from the body, though that does not matter in this case). At large distances from the body, it appears as a gravitational point source of mass $M^A$.
\item The \textbf{passive (gravitational) mass} $M^P$ is defined by
\be
\vctr{F}=M^P \vctr{g}
\ee
where $\vctr{F}$ is the total gravitational force on the body and $\vctr{g}$ is an external gravitational field that can be idealized as constant across the extent of the body. With respect to fields like this, the body responds to gravity as if it were a point particle of mass $M^P$.
\item The \textbf{inertial mass} $M^I$ is defined by 
\be
\vctr{F}=M^I \vctr{a}
\ee
where $\vctr{F}$ is the total external force on the body and $\vctr{a}$ is the acceleration of its center of mass. 
\item The \textbf{integrated mass density} $M^{\mathrm{int}}$ is just the spatial integral of the mass density over the extent of the body,
\be
M^{\mathrm{int}}=\int \rho.
\ee
Integrating \eqref{eq:contgeneral-mass} over the extent of the body shows that it is conserved; for a body that consists of a swarm of particles, it is just the sum of their individual masses.
\end{itemize}
Active and passive mass need to be equal\footnote{Strictly speaking we just need $M^A\propto M^P$ and $M^P\propto M^I$; equality is then a conventional choice in each case.} in order for overall momentum to be conserved (this is the same argument we used previously to establish that gravity is sourced by mass). Passive mass and inertial mass need to be equal in order for the Strong Equivalence Principle to hold: if $M^P/M^I$ varied from one body to another, those bodies would accelerate differently in the same gravitational field. For Newtonian gravity, the linearity of the field equations makes it immediate that all four definitions coincide; later we will consider nonlinear generalizations of Newtonian gravity where this becomes a substantive requirement.

\section{Generalizing inertial forces: rotation}\label{generalizinginertial}

The basic logic of the previous section was:
\begin{enumerate}
\item Introduce an inertial force by transforming to a non-inertial coordinate system.
\item Allow that force to vary smoothly from spacetime point to spacetime point, so that locally it can be interpreted as inertial but globally it cannot (and so can be taken as localizing the global inertial structure represented by the inertial force).
\item Seek a dynamical system where the locally inertial force is determined by the matter distribution, given the requirement that empty space in the transformed coordinates satisfies the dynamical equations and other reasonable assumptions.
\end{enumerate}
That points to an extremely natural generalization: instead of considering (just) the rectilinearly accelerating coordinate systems obtained by time-dependent translations, we can consider a wider class of non-inertial coordinate systems, represent a larger part of the inertial structure of Galilean spacetime via inertial forces, localize \emph{those} forces, and seek dynamical equations for them. A natural starting point is \emph{rotational motion}.

\subsection{Physics in rotating frames}

 As is well known,  the equation of motion for an inertially moving particle in a rotating reference frame is
\be
m \dbd{\vctr{v}}{t}=-m \vctr{\omega}\times(\vctr{\omega}\times \vctr{x}) + 2 m \vctr{\omega}\times\vctr{v}+m\partial_t\vctr{\omega}\times \vctr{x}, 
\ee
where $\vctr{x}$ and $\vctr{v}$ are respectively the particle's position and velocity, and  $\vctr{\omega}$ is the angular velocity of the coordinate transformation (so that the frame has angular velocity $-\vctr{\omega}$). We can represent this as an inertial force
\be
\vctr{F}=m (\vctr{G}+ \vctr{v}\times \vctr{H})
\ee
where
\be\label{G-for-rotations}
\vctr{G}=-\vctr{\omega}\times(\vctr{\omega}\times \vctr{x}) + \partial_t\vctr{\omega}\times \vctr{x}
\ee
is a universal force field of the kind we have already encountered, which from here on I call a \emph{translational field}, and 
\be\label{H-is-omega}
\vctr{H}=-2 \vctr{\omega}
\ee
is the \emph{Coriolis field} (see, e.g., \citep{costanatariocoriolis}). So one natural generalization of Newtonian gravity would be to generalize our notion of universal forces to include Coriolis forces, and to seek a way for matter to source those forces too.

\subsection{Localizing the Coriolis field}\label{localizing-coriolis}

Following the same schema we used for Newtonian gravity, we should first seek field equations for $\vctr{G}$ and $\vctr{H}$ that are satisfied in empty space, and by taking the divergence and curl of \eqref{G-for-rotations}--\eqref{H-is-omega}, we can determine that these equations must be
\begin{subequations}
\begin{align}\label{freefield-rotation}
\vctr{\nabla} \cdot \vctr{G}&=\frac{1}{2}H^2 \\
\vctr{\nabla} \times \vctr{G}&=-\partial_t \vctr{H}\\
\vctr{\nabla} \cdot \vctr{H}&=0\\
\vctr{\nabla} \times \vctr{H}&=0
\end{align}
\end{subequations}
and can then ask how these equations can be modified in the presence of matter.

Writing $\rho$ and $\vctr{J}$ for the mass density and momentum density of a distribution of matter, the gravitational force density on that matter is
\be
\vctr{f}=\rho \vctr{G}+\vctr{J}\times \vctr{H}.
\ee
That is: the translational field couples linearly to the mass density and the Coriolis field couples linearly to the momentum density. This coupling is again universal and fixed in form by the requirement that $\vctr{G}$ and $\vctr{H}$ can be interpreted locally as inertial fields. Momentum-conservation considerations then suggest that the fields should be sourced linearly \emph{by} $\rho$ and $\vctr{J}$. Since $\rho$ is a scalar and $\vctr{J}$ is a vector, that pretty much fixes the form of the field equations to be
\begin{subequations}
\begin{align}
\vctr{\nabla} \cdot \vctr{G}&=-4\pi G_N \rho + \frac{1}{2}H^2 \\
\vctr{\nabla} \times \vctr{G}&=-\partial_t \vctr{H}\\
\vctr{\nabla} \cdot \vctr{H}&=0\\
\vctr{\nabla} \times \vctr{H}&=-4\pi K G_N \vctr{J}
\end{align}
\end{subequations}
with $K$ a new constant of nature (expressed without loss of generality as a multiple of $G_N$) whose sign and magnitude are undetermined. (The two sourceless equations are a pseudovector and a pseudoscalar, and there are no available sources with these transformation properties.) Under these field equations, the translational force from a stationary point particle would be given by the usual inverse-square form, and the Coriolis field from a moving chunk of matter\footnote{As we will shortly see, this theory does not actually permit an isolated moving particle to generate a Coriolis field; this result must be understood as the field due to some part of a larger distribution of moving masses.} of mass $m$ and velocity $\vctr{v}$ would have the magnetic form
\be
\vctr{H}(\vctr{r}) = -K G_N m\frac{\vctr{v}\times \vctr{r}}{r^3}
\ee
evaluated at vector displacement $\vctr{r}$ from the particle's location.

This expression, however, shows us that we need to be a little cautious with appeals to momentum conservation as a constraint, since in general this interaction does \emph{not} conserve mechanical momentum: consider particles of masses $m_1,m_2$ and velocities $\vctr{v}_1,\vctr{v}_2$, and let $\vctr{r}$ be the vector from the first to the second; then the net force on the two particles is
\begin{align}
\vctr{F}&=-G_N K \frac{m_1m_2}{r^3}(\vctr{v}_2\times (\vctr{v}_1 \times \vctr{r}) - \vctr{v}_1 \times (\vctr{v}_2\times \vctr{r}))\nonumber \\
&=-G_N K \frac{m_1m_2}{r^3}\vctr{r}\times(\vctr{v}_1\times \vctr{v}_2)
\end{align}
which is generically nonzero. The only way to preserve momentum conservation is for the Coriolis field itself to carry momentum. (This is a gravitomagnetic analog of the familiar electromagnetic result that magnetic interactions violate Newton's Third Law.)

We can, however, consider the static case, where the overall distribution of mass and momentum of the matter is unchanging. For instance, we could consider the gravitational interactions of two rotating rings of matter, held in place by nongravitational forces. In this case, the gravitational fields are time-independent and momentum conservation reduces to Newton's Third Law for the matter distribution. 

Working in the continuum regime and to linear order, the net gravitational force on a distribution of matter with mass density $\rho$ and momentum density $\vctr{J}$ is then
\begin{align}
\vctr{F}=-G_N\int\dr{^3x}\dr{^3y} \left\{\rho(\vctr{x})\rho(\vctr{y})\frac{\vctr{r}}{r^3}+K\vctr{J}(\vctr{x})\times\left(\vctr{J}(\vctr{y})\times \frac{\vctr{r}}{r^3}\right)\right\}
\end{align}
where $\vctr{r}=\vctr{x}-\vctr{y}$, and imposing momentum conservation in the static case amounts to requiring that $\vctr{F}=0$. Expanding the double cross product in the second term, and using the fact that the integral over $\vctr{x}$ and $\vctr{y}$ of a function odd under interchange of $\vctr{x}$ and $\vctr{y}$ must vanish, this simplifies to
\be
0=\int\dr{^3y} \vctr{J}(\vctr{y}) \int\dr{^3x}\vctr{J}(\vctr{x})\cdot  \frac{\vctr{r}}{r^3}.
\ee
Writing $\vctr{r}/r^3=-\vctr{\nabla}_{\vctr{x}}(1/r)$ and integrating by parts, it reduces further to
\be
0=\int\dr{^3y} \vctr{J}(\vctr{y}) \int\dr{^3x}\frac{1}{r}\vctr{\nabla}_{\vctr{x}}\cdot \vctr{J}(\vctr{x})
\ee
which vanishes in the static case where $\vctr{\nabla}\cdot \vctr{J}=0$.

To see that universality is again required, specialize to two rotating rings. If the Coriolis force depended on the composition of a ring and not just on its momentum current, then by varying the composition of one ring we would change the force it exerted \emph{on} the other ring while leaving unchanged the force exerted on it \emph{by} the other ring, and momentum conservation would be violated. The basic structure is the same as in the Newtonian case: since the response of matter to gravitational forces is universal, the generation of gravitational forces \emph{by} matter is likewise constrained to be universal.

\subsection{Problems with mass conservation}

Unfortunately, this set of equations has a fatal flaw. Taking the divergence of both sides of the fourth equation yields $-4\pi G_N K \vctr{\nabla} \cdot \vctr{J}=0$, which together with mass conservation entails $K \partial_t \rho=0$ --- that is, the equations are incompatible with time-varying mass distributions unless $K=0$. In parallel with the case of the magnetostatic bootstrap, we  might be tempted to solve this by replacing the fourth equation with
\be
\vctr{\nabla} \times \vctr{H}=-4\pi K G_N \vctr{J}+ K \partial_t\vctr{G}
\ee
but if we do so, the free-field equations will no longer be solved in arbitrary accelerating coordinate systems --- in other words, this extra term breaks the interpretation of $\vctr{G}$ as a localization of inertial structure.\footnote{A theory of this kind was in fact proposed by Heaviside~\citep{heaviside1893analogy}, building on an initial proposal by Maxwell~\citep{maxwell1865dynamical} (see \citep{mcdonald1997why} for further discussion), but it made no use of the Equivalence Principle.}

One fix would be to set $K=0$, giving up on the idea that the Coriolis field can be sourced by matter. The resultant field equations are 
\begin{subequations}
\begin{align}
\vctr{\nabla} \cdot \vctr{G}&=-4\pi G_N \rho + \frac{1}{2}H^2 \\
\vctr{\nabla} \times \vctr{G}&=-\partial_t \vctr{H}\\
\vctr{\nabla} \cdot \vctr{H}&=0\\
\vctr{\nabla} \times \vctr{H}&=0
\end{align}
\end{subequations}
and in fact this theory is well known: it is the \emph{Newton--Cartan theory} (see \citep{costanatariocoriolis} for a presentation in this form). If supplemented by appropriate boundary conditions it reduces to Newtonian gravity in a rotating reference frame; it is compatible with more general (divergence-free, curl-free) Coriolis fields, but these fields cannot be sourced by matter.

There is another alternative, though: we could consider further localization of the spacetime structure of Newtonian physics.

\section{Generalizing inertial forces: spatial scale}\label{localizedscale}

Newtonian gravitation is invariant under simultaneous rescaling of length, mass, and time, which suggests that the overall scale of space is dynamical in character and might also be thought of as part of the inertial structure. This in turn suggests considering the larger class of \emph{conformal transformations} of Euclidean space, in which the overall spatial scale is also allowed to vary. 

\subsection{Physics in conformally transformed frames}\label{physics-in-conformal-frames}

A conformal transformation is a transformation $\vctr{x}\rightarrow \tilde{\vctr{x}}$ that satisfies
\be
\pbp{\tilde{x}_i}{x_j}=R_{ij}(\vctr{x})S(\vctr{x})
\ee
for some possibly spacetime-dependent rotation matrix $R_{ij}(\vctr{x})$ and scalar $S$, so that it preserves angles but not necessarily lengths. Rotations and translations are conformal transformations; so are dilations (uniform stretchings of space), and so are so-called \emph{special conformal transformations} (see, e.g.,~\citep[p.~97]{difrancesco1997}), which take the form
\be
\tilde{\vctr{x}}=\frac{\vctr{x}+x^2 \vctr{b}}{1+2\vctr{b}\cdot\vctr{x}+b^2 x^2}
\ee
for arbitrary vector $\vctr{b}$. There are thus 3 independent special conformal transformations; they, the dilations, rotations, and translations jointly generate the 10-dimensional \emph{conformal group} of 3-dimensional Euclidean space. (The special conformal transformations are singular where the denominator vanishes, but since in any case we are interested in forces that \emph{locally} look inertial, this need not concern us.) 

Under a (time-independent) conformal transformation, an ordinary real function transforms like $f\rightarrow \tilde{f}$, where
\be
\tilde{f}(\tilde{\vctr{x}})= f(\vctr{x}),
\ee
and an ordinary vector field transforms like $\vctr{V} \rightarrow \tilde{\vctr{V}}$, where
\be
\tilde{\vctr{V}}(\tilde{\vctr{x}})=\vctr{R}(\vctr{x})S(\vctr{x})\vctr{V}(\vctr{x})
\ee
or equivalently
\be
\tilde{V}_i(\tilde{\vctr{x}})=R_{ij}(\vctr{x})S(\vctr{x})V_j(\vctr{x})
\ee

However, it will be useful to consider vector-like and scalar-like objects with more general transformation properties. An object represented in a given coordinate system by a scalar function $f$ has \emph{conformal weight $N$} (equivalently: is a conformal-weight-$N$ scalar) if its transformation law is $f\rightarrow \tilde{f}$ with
\be
\tilde{f}(\tilde{\vctr{x}})=S^N(\vctr{x}) f(\vctr{x}).
\ee
And an object represented by a vector function $\vctr{V}$ has conformal weight $N$ (equivalently: is a conformal-weight-$N$ vector field) if its transformation law is $\vctr{V} \rightarrow \tilde{\vctr{V}}$ with
\be
\tilde{\vctr{V}}(\tilde{\vctr{x}})=\vctr{R}(\vctr{x})S^{N+1}(\vctr{x})\vctr{V}(\vctr{x})
\ee
With these conventions, ordinary vectors and scalars have conformal weight zero. (Be warned that several different conventions for conformal weight are used in the literature.) We can also define \emph{pseudoscalars} and \emph{pseudovectors} of arbitrary conformal weight, which behave oppositely to ordinary scalars/vectors under reflections. The cross product of vector fields of conformal weight $M,N$ is a pseudovector of conformal weight ($M+N+1$), and the dot product of those vector fields is a scalar of conformal weight $(N+M+2)$. And while we won't really use it, a 2nd rank tensor of conformal weight $N$ transforms like
\be
\tilde{\vctr{T}}(\tilde{\vctr{x}})=\vctr{R}(\vctr{x})\vctr{T}(\vctr{x})\vctr{R}^T(\vctr{x})S^{N+2}(\vctr{x})
\ee

Under a (time-independent) conformal transformation, the equation of motion of an inertially moving particle takes the form
\be
m \dbd{\vctr{v}}{t}=2m (\vctr{\nabla} \ln S \cdot \vctr{v})\vctr{v} - m v^2\vctr{\nabla} \ln S.
\ee
Note that conformal transformations are not in general dynamical symmetries, so even time-independent conformal transformations lead to inertial forces unless $S$ is spatially constant. In fact it will be convenient to represent this in a different way: we can rewrite it as
\be
\ddt\left(\frac{m}{S^2} \vctr{v}\right) = - \left(\frac{m}{S^2}\right)v^2 \frac{\vctr{\nabla}S}{S}.
\ee
This combines an inertial force field $\vctr{\nabla}\ln S=\vctr{\nabla}S/S$ which couples to a particle's squared velocity, and an \emph{inertial mass dilation}: the inertial mass of the particle is its mass in a frame where $S=1$, divided by the square of $S$.

When we generalize to the time-dependent case, we will find that $\vctr{F}$ and $\Omega$ transform the same way as before, but that $\vctr{G}$ and $\vctr{H}$ pick up additional terms depending on the time derivatives of the transformation; geometrically, this indicates that these inertial fields are not purely spatial. Nonetheless I will still use the `conformal weight' terminology to define them, and will continue to write $\tilde{\vctr{G}}$, $\tilde{\vctr{H}}$ for their transformations under the time-independent part of $\vctr{x}\rightarrow \tilde{\vctr{x}}$.

A time-dependent conformal transformation (specified by time-dependent $\vctr{R}(t), S(t)$) can usefully be characterized by the vector field
\be\label{eq:conformalflow}
\vctr{u} = (\vctr{R} S )^{-1}\partial_t\tilde{\vctr{x}}
\ee
which I refer to  as the \emph{flow} of the conformal transformation. We can also define the objects
\begin{subequations}\label{eq:conformalobjects}
\begin{align}
\sigma &= S^{-1}\partial_t S\\
\omega_{ij}&=[\vctr{R}^{-1}\partial_t \vctr{R}]_{ij}\label{vorticity}\\
(\vctr{\omega}\times \vctr{v})_i &\equiv \omega_{ij}v_j
\end{align}
\end{subequations}

For future purposes it will be helpful to introduce a scalar field $\Omega$ which locally represents conformal structure. If we are just doing inertial physics in a conformally transformed frame then $\Omega=S$, but shortly we will consider more general $\Omega$ that only locally represent a coordinate change. 
The general equation  for the rate of change of momentum of a  moving particle in a conformally transformed reference frame is then
\be\label{conformalequation}
\ddt\left(\frac{m}{\Omega^2} \vctr{v}\right)  = \frac{1}{\Omega^2}\vctr{F}+\left(\frac{m}{\Omega^2}\right)(\vctr{G}+\vctr{v}\times \vctr{H} - v^2 \vctr{C})
\ee
with  $\vctr{C}=\vctr{\nabla} \ln \Omega$. The key to proving this result is:
\begin{theorem}
\label{thm1}
The equation of motion \eqref{conformalequation} is covariant under the group \textbf{TDConf} of time-dependent conformal transformations provided that $m$ is independent of the particle's velocity. If this condition holds, the fields $(\vctr{F},\vctr{G},\vctr{H},\Omega)$ have conformal weights $(0,0,-1,1)$ and $\vctr{G}$, $\vctr{H}$ additionally transform as follows at a reference time  under a time-dependent transformation that is the identity at that time:
\begin{subequations}\label{eq:conf-G-H-trans}
\begin{align}
\vctr{H}\rightarrow \vctr{H}'&=\vctr{H}-2\vctr{\omega}-2 \vctr{u}\times\vctr{C}\\
\vctr{G}\rightarrow\vctr{G}'&= \vctr{G}- \vctr{u}\times \vctr{H}'-2\vctr{u}\partial_t \ln\Omega+\partial_t \vctr{u}+\vctr{\omega}\times\vctr{u}-\sigma\vctr{u}+u^2 \vctr{C}.\label{generaltransformation}
\end{align}
\end{subequations}
The action of an arbitrary time-dependent conformal transformation on the fields is then expressible as the conjunction of these two transformations.
\end{theorem}
\begin{proof}See Appendix~\ref{proof-of-theorem-1}.\end{proof}
Given this theorem, we can simply apply a time-dependent conformal transformation to flat inertial-coordinatized spacetime to obtain \eqref{conformalequation}.
It is worth noting that although the various inertial force \emph{fields} transform with various conformal weights, the \emph{forces}, whether inertial or non-inertial, all have the same weight. This reflects the fact that it is to some degree a matter of convention whether we take a universal force to be inertial or not. Indeed, in applications of Newtonian gravity it often makes sense to treat the gravitational effects of distant bodies as inertial (and then transform them away) while treating the forces between nearby bodies as `ordinary' forces --- as we do in, for instance, analyzing the Earth--Moon system or the moons of Jupiter.

Equation~\eqref{conformalequation}, together with the invariance of unrescaled mass, comprise the equations of motion for a point particle. However, since we seek continuum field equations it will also be useful to have the continuum versions, which are:
\begin{subequations}
\begin{align}
\label{eq:contforce}\partial_t \vctr{J}_R + \vctr{\nabla} \cdot \vctr{T}_R&=\vctr{f}_R+\rho_R \vctr{G}+\vctr{J}_R\times \vctr{H} -(\tr\vctr{T}_R)  \vctr{C} \\
\label{eq:contmass}\partial_t \rho + \vctr{\nabla} \cdot \vctr{J}&=0,
\end{align}
\end{subequations}
where $\vctr{f}$ is the force density (the force per unit volume) due to non-inertial forces, and $X_R=X/\Omega^2$ denotes a conformally rescaled quantity.

These equations of motion will be the starting point for our further generalization of Newtonian gravity, which we can call \emph{conformal gravity}: we will try to treat $\vctr{G}$, $\vctr{H}$, and $\Omega$ as general spacetime-dependent fields, locally but not globally interpretable as inertial forces, and seek to couple them to matter. 
We will see that localizing the full conformal structure is exactly what is required to cure the deficiencies of our previous theory: an appropriate choice of the relative strengths of the coupling of matter to $\vctr{H}$ and $\vctr{C}$ will ensure that mass conservation is preserved even if the mass distribution is time-dependent.

\subsection{Field equations in empty space}\label{equations-in-empty-space}

As with the first two theories, our first step is: determine the fields which describe empty flat spacetime in transformed coordinates, and then find local field equations that are solved by those fields. The process is much more involved for general conformal transformations, and I leave the details to the Appendix, but the result is: 
\begin{theorem}\label{thm2}
The equations
\begin{subequations}\label{vacuumfieldequations}
\begin{align}
\Omega^{1/2}\nabla^2 (\Omega^{-1/2})&=0 \\
\Omega^2\vctr{\nabla}\cdot(\Omega^{-2}\vctr{H})&=0 \\
\Omega \vctr{\nabla}\times(\Omega^{-1}\vctr{H})&=-4 \partial_t \vctr{C}\\
\Omega^3\vctr{\nabla}\cdot (\Omega^{-3}\vctr{G})&=3\partial_t^2\ln\Omega-3(\partial_t\ln\Omega)^2+\frac{1}{2}H^2\\
\Omega^2\vctr{\nabla}\times(\Omega^{-2}\vctr{G})&=-\partial_t\vctr{H}+2 \vctr{H}\partial_t\ln\Omega
\end{align}
\end{subequations}
are satisfied for all inertial field triples $(\vctr{G},\vctr{H},\Omega)$ describing conformal transformations from non-accelerated and non-rotated Cartesian coordinates. 
\end{theorem}
\begin{proof}See Appendix~\ref{proof-of-theorem-2}.\end{proof}

With suitable boundary conditions, and if they are imposed everywhere in space, the equations will characterize exactly the inertial forces. 
But we are not interested in imposing those restrictions: we want to couple the fields to matter, which will involve adding source terms to the equations.

\subsection{Single-particle fields and the velocity scale}\label{oneparticle}

Again we consider a static distribution of matter: concretely, let's consider two compact bodies  (two rocks, say, or two gas clouds) with no net momentum density and at large distance from one another. The vacuum field equations are nonlinear, but in the limit of large separation we can assume that nonlinear terms can be neglected so that we can meaningfully speak of the fields generated by each chunk separately. In the absence of any other preferred direction, these fields will have to act along the line between the two bodies; at large distances, the vacuum field equations will require these forces to have inverse-square form.

If we define $M_n$ and $\Sigma_n$, respectively, as the integral of the mass density and of the trace of the stress tensor over the $n$th body, and $\vctr{C}_n$ and $\vctr{G}_n$ as the fields due to that body, momentum conservation requires
\be
(M_2 \vctr{G}_1 - \Sigma_2 \vctr{C}_1)=- (M_1 \vctr{G}_2 - \Sigma_1 \vctr{C}_2)
\ee
which can be satisfied only if
\begin{subequations}
\begin{align}
\vctr{G}_n&=-\frac{G_N  \hat{\vctr{r}}}{r^2}(M_n + D \Sigma_n)\\
\vctr{C}_n&=+\frac{G_N \hat{\vctr{r}}}{r^2}(DM_n + E \Sigma_n)
\end{align}
\end{subequations}
where $\vctr{r}$ is the vector from particle $n$ to the other particle and where $D$ and $E$ are new constants of nature.

Since the non-Coriolis forces balance, so must the Coriolis forces, and the arguments of section~\ref{localizing-coriolis} fix those forces too. Written in terms of point particles, where $\Sigma=m v^2$, the (large-distance, linear-approximation) force fields are then given by
\begin{subequations}
\begin{align}
\vctr{G}(m,\vctr{v},\vctr{r})&=-\frac{mG_N \hat{\vctr{r}}}{r^2}(1 + v^2 D) +O(1/r^3)\\
\vctr{H}(m,\vctr{v},\vctr{r})&=-\frac{mG_N}{r^2} B \vctr{v}\times\hat{\vctr{r}}+O(1/r^3) \\
\vctr{C}(m,\vctr{v},\vctr{r})&=+\frac{m G_N  \hat{\vctr{r}}}{r^2} (D + v^2 E) +O(1/r^3)
\end{align}
\end{subequations}
where $B,D,E$ are unknown constants that parametrize the  strengths of the various field components relative to the Newtonian gravitational force field. 

Let's consider the dimensions of these constants. Observe firstly that the Coriolis force between two particles scales with $1/r^2$ times the product of their masses times the cross product of their velocities, while the Newtonian force between them scales with $1/r^2$ times the product of their masses. Mathematically the ratio of those forces is $B |\vctr{v}_1\times \vctr{v}_2|$ and it follows that $B$ must have the dimensions of inverse velocity squared. Similarly, $D$ must have those dimensions. In other words, if any of the non-Newtonian parts of the gravitational field actually exist, there must be a new dimensionful constant in addition to $G_N$, and it must have the dimensions of velocity.\footnote{Obviously it's conventional whether we take it to have the dimensions of velocity, velocity-squared, 1/velocity, etc.} Picking a letter from the alphabet entirely at random, let us call this unknown constant $c$. By definition $D=1/c^2$, and we introduce a dimensionless constant $\beta$ such that $B=\beta/c^2$ (so that $\beta$ fixes the ratio of the Coriolis field to the conformal-gradient field). Similarly, $E$ must have the dimensions of inverse velocity to the fourth power, and we can write $E\equiv \lambda/c^4$ with $\lambda$ dimensionless.

The leading-order form of a single particle's fields is then
\begin{subequations}
\begin{align}
\vctr{G}(m,\vctr{v},\vctr{r})&=-\frac{mG_N \hat{\vctr{r}}}{r^2}(1 + v^2/c^2) +O(1/r^3)\\
\vctr{H}(m,\vctr{v},\vctr{r})&=-\frac{mG_N }{r^2} (\beta/c^2) \vctr{v}\times\hat{\vctr{r}}+O(1/r^3) \\
\vctr{C}(m,\vctr{v},\vctr{r})&=+\frac{mG_N\hat{\vctr{r}}}{r^2c^2} (1 + \lambda v^2/c^2)+O(1/r^3).
\end{align}
\end{subequations}

This characterizes the (large-distance, linear-approximation) forces uniquely up to the unknown velocity $c$ and the dimensionless constants $\beta,\lambda$. There is a subtlety, however: as written, assuming $c$ is real-valued then the conformal-gradient force is attractive, which does not in fact follow from our analysis. This results because the actual dimensionful constant is $c^2$, which can be positive or negative. Formally, we can include the possibility of a repulsive conformal-gradient field by allowing $c$ to be pure imaginary. Ultimately, we will find that it is an empirical matter whether the field is attractive or repulsive; for expository convenience I will write as if $c$ is real, but all the mathematical results should follow in either case.

In any case, we can see from these expressions that the size of non-Newtonian gravitational effects (compared to Newtonian effects) is controlled by two dimensionless quantities: $v^2/c^2$, and $(G_NM/rc^2)$. Furthermore, the success of Newtonian gravity at modeling the Solar System tells us that $c$ must be large enough that these quantities are $\ll 1$ for Solar System objects.

Now that we have identified this scale, we lower our sights a little. Instead of seeking an \emph{exact} theory which treats rotational inertia and scale as localized and matter-dependent, we shall (at least for the moment) content ourselves with finding a theory that is valid to the \emph{first post-Newtonian approximation} (1PN approximation): that is, a theory in which terms of order $(v^2/c^2)$ and $(G_NM/rc^2)$ cannot be neglected, but $(v^2/c^2)^2$ and $(G_NM/rc^2)^2$ and the like can be neglected. (The \emph{Newtonian approximation}, of course, consists of neglecting even these terms, and is just Newtonian gravity.) This is to adopt an effective-field-theory approach to our post-Newtonian gravitational theory, and will allow us to simplify our task considerably. As an initial step in that simplification, the $v^2$ source for $\vctr{C}$ is proportional to  $c^{-4}$ and so at 1PN we can discard this term, leaving $\beta$ as the only dimensionless parameter. 

It is useful to identify two aspects of the 1PN approximation. The \emph{low-velocity approximation} is that matter (at least that matter which generates gravitational fields) has velocity satisfying $v^2/c^2\ll 1$. The \emph{weak-field approximation}, again generalizing to a multiparticle distribution, is that $\Phi(\vctr{x})/c^2\ll 1$ whenever matter is located at $\vctr{x}$.
They are logically independent; however, the virial theorem of celestial mechanics (reviewed in Appendix~\ref{appendix-masses}) tells us that $v^2\sim \Phi$, so for a group of bodies moving predominantly under their own gravitational interactions we can expect them to coincide. I will denote a term as `$O(1/c^2)$' if at leading order it depends linearly on $v^2/c^2$ and/or $\Phi/c^2$, as `$O(1/c^4)$' if it depends quadratically on these terms, and so on. The Newtonian regime is then obtained by neglecting terms of $O(1/c^2)$ or higher; the 1PN regime, by neglecting terms of $O(1/c^4)$ or higher. Our analysis so far shows that $\vctr{H}=O(1/c^2)$ and $\Omega=1+O(1/c^2)$; thus, the Newtonian regime corresponds to setting $\Omega=1$ and discarding all terms in $\vctr{H}$ and $\vctr{C}$; the 1PN regime corresponds to discarding terms quadratic in $\vctr{H}$ and $\vctr{C}$ and setting $\Omega=1$ whenever it multiplies any such term.

One further consequence of the 1PN approximation is that it restricts the size of the rotation and conformal transformations that we can consistently consider. Since a Coriolis field is generated by a time-dependent rotation, and a conformal field by a time-independent dilation or special conformal transformation, we need to assume that the parameters of any such transformations are $O(1/c^2)$. (Time-independent rotations and time-dependent translations are not so restricted.) In terms of the parameters $\vctr{u},\vctr{\omega},\sigma$ from \eqref{eq:conformalflow}--\eqref{eq:conformalobjects}, this entails taking the non-spatially-constant part of $\vctr{u}$, and hence $\sigma$ and $\vctr{\omega}$, to be $O(1/c^2)$.

To avoid notational clutter, from here on I will assume that we are in the 1PN regime and will silently discard terms of $O(1/c^4)$ or higher.

\subsection{Field equations for conformal gravity}\label{field-equations-conformal}

The 1PN approximation significantly simplifies the empty-space field equations~\eqref{vacuumfieldequations} for conformal gravity, to 
\begin{subequations}\label{vacuumfieldequations-1pn}
\begin{align}
\vctr{\nabla}\cdot \vctr{C}&=0 \\
\vctr{\nabla}\cdot \vctr{H}&=0 \\
\vctr{\nabla}\times\vctr{H}&=-4 \partial_t \vctr{C}\\
\Omega^3\vctr{\nabla}\cdot (\Omega^{-3}\vctr{G})&=3\partial_t^2 \Omega\\
\Omega^2 \vctr{\nabla}\times(\Omega^{-2}\vctr{G})&=-\partial_t\vctr{H}.
\end{align}
\end{subequations}
Since in the continuum equations of motion \eqref{eq:contforce}--\eqref{eq:contmass} matter only couples to gravity through the (conformally rescaled) mass density, momentum density, and stress trace, momentum conservation in the static case means that it can only be sourced \emph{by} these quantities. Dimensional analysis and the leading-order Newtonian result then tell us that to leading order the sources must be $G_N\rho_R$, $(G_N/c^2)\vctr{J}_R$, and $(G_N/c^2)\tr \vctr{T}_R$. (Indeed, we could have deduced the need for the velocity parameter $c$ directly by applying dimensional analysis to these quantities, albeit at the cost of less physical understanding.)

There is one complication: nothing in the leading-order approximation prohibits any of the sources being multiplied in the field equations by some power of $\Omega$, since this generates nonlinear effects that are invisible to the linear-order approximation of section~\ref{oneparticle}. This has no effect at 1PN for the coupling to $\vctr{J}$ or $\vctr{T}$, but it could affect coupling to mass density. To allow for this, let us define the rescaled density
\be
\rho_*=\Omega^q \rho = \Omega^{q+2}\rho_R
\ee
for some unknown parameter $q$. 

This, together with the analysis of the point-particle gravitational field in section~\ref{oneparticle}, forces the form of the field equations (at 1PN) to be
\begin{subequations}
\begin{align}
\vctr{\nabla}\cdot \vctr{C}&=(4\pi G_N /c^2) \rho_* \\
\vctr{\nabla}\times\vctr{H}+4 \partial_t \vctr{C}&=-(4\pi G_N/c^2)\beta \vctr{J}\\
\vctr{\nabla}\cdot \vctr{G}-3\vctr{C}\cdot\vctr{G}- 3\partial_t^2 \Omega&=-4\pi G_N\left( \rho_* +\frac{\tr \vctr{T}}{c^2}\right) 
\end{align}
\end{subequations}
with the other two equations remaining sourceless, where I have reintroduced the dimensionless parameter $\beta$ from section~\ref{oneparticle} (and where I haven't bothered to distinguish what rescaling is used for the momentum current $\vctr{J}$ and the stress tensor $\vctr{T}$, since at 1PN it has no effect).
This is pretty much the most general way we could couple our sources to the fields at 1PN order, given that action-reaction considerations require the dimensionless parameters in $\vctr{\nabla} \cdot \vctr{C}$ and $\vctr{\nabla} \cdot \vctr{G}$ to match: the other two equations have the wrong parity to couple to matter.

In fact, $\beta$ is uniquely determined. If we take the divergence of the third equation and substitute in the first, we get
\be
(4\pi G_N/c^2)(4\partial_t \rho_*+\beta \vctr{\nabla}\cdot \vctr{J})=0.
\ee
But from \eqref{eq:contmass}, $\partial_t \rho_*+\vctr{\nabla}\cdot\vctr{J}=O(1/c^2)$, so consistency requires $\beta=4$ (and that the Coriolis and conformal-gradient interactions are either both attractive or both repulsive), and the complete set of field equations are
\begin{subequations}\label{fullfieldequations}
\begin{align}
\vctr{\nabla}\cdot \vctr{C}&=(4\pi G_N /c^2)\rho_* \label{eq:sec5-1}\\
\vctr{\nabla}\cdot \vctr{H}&=0 \label{eq:sec5-2} \\
\vctr{\nabla}\times\vctr{H}+4 \partial_t \vctr{C}&=-4(4\pi G_N /c^2)  \vctr{J} \label{eq:sec5-3}\\
\vctr{\nabla}\cdot \vctr{G}-3\vctr{C}\cdot\vctr{G}- 3\partial_t^2 \Omega&=-4\pi G_N \left(\rho_* +\frac{\tr \vctr{T}}{c^2}\right)  \label{eq:sec5-4} \\
\vctr{\nabla}\times\vctr{G}-2 \vctr{C}\times \vctr{G}+\partial_t\vctr{H}&=0. \label{eq:sec5-5}
\end{align}
\end{subequations}
We can see now how the localization of the conformal gradient fixes the Coriolis-field theory of section~\ref{generalizinginertial}: a specific ratio of conformal-gradient coupling to Coriolis coupling is required to ensure mass conservation. This is ultimately the justification for localizing the conformal factor: without it, we would not have been able to write a consistent theory.

Let's pause briefly to consider boundary conditions for conformal gravity: we can distinguish three, decreasingly stringent, conditions.
\begin{itemize}
\item ``Island-universe'' boundary conditions consist of requiring $(\Omega-1)$, $\vctr{G}$, and $\vctr{H}$ all to tend to zero at infinity. They describe an isolated system, far from any other sources of gravitation. These boundary conditions imply that $\vctr{C}+\vctr{G}/c^2=0$ everywhere, since both fields are just inverse-square fields generated by the mass distribution. They also imply that $\Omega$ has the explicit form
\be
\Omega=1-\frac{1}{c^2}\Phi
\ee
where $\Phi$ is the $1/r$ potential \eqref{phi-explicit}.
\item A subsystem in a larger island-universe system will need boundary conditions that describe the background fields for that system, which need not vanish. However, the condition $\vctr{C}+\vctr{G}/c^2=0$ will still hold for those background fields, so the subsystem's boundary conditions will still enforce $\vctr{C}+\vctr{G}/c^2=0$. We can call these boundary conditions ``island-subsystem'' boundary conditions. 
\item Nothing\footnote{Well, provided that our theory is actually covariant under these transformations --- and, as we will shortly see, this is not entirely correct. But the discussion carries over to the theory I discuss in section~\ref{temporalscale}, which is covariant.} stops us from describing a subsystem's physics in an accelerating or conformally transformed frame, and in those frames, new inertial forces are introduced and generically $\vctr{C}+\vctr{G}/c^2\neq 0$. But even in those frames, the physically sourced part of $\vctr{G}$ and $\vctr{C}$, which generates the spatial variation of $\vctr{G}$ and (to 1PN) of $\vctr{C}$, has the same source. So given any set of boundary conditions which idealize the effect of physical sources, the condition $\partial_i(C_j+G_j/c^2)=0$ should hold. Call any boundary condition that enforces this requirement \emph{source-consistent}.
\end{itemize}
I will assume for the rest of the paper that any boundary conditions are at least source-consistent.

\subsection{Determining the source}

Section~\ref{oneparticle} established momentum conservation only at linear order. It will turn out that momentum conservation in the full nonlinear theory (at 1PN) fixes the parameter $q$. Again we consider a static distribution of matter. Coriolis effects are linear and so the previous analysis already establishes momentum conservation there, so we only need to look at the remaining equations. Using $\Omega=1-\Phi/c^2$ and $\vctr{C}=-\vctr{\nabla} \Phi/c^2$, this amounts to establishing the consistency of:
\begin{subequations}
\begin{align}
\vctr{\nabla} \cdot \vctr{G}+\frac{3}{c^2}\vctr{G}\cdot \vctr{\nabla} \Phi&=-4\pi G_N \left(\rho(1-\frac{q}{c^2}\Phi)+\frac{1}{c^2}\tr\vctr{T} \right)\\
\vctr{\nabla} \times \vctr{G}-\frac{2}{c^2}\vctr{G}\times \vctr{\nabla} \Phi &=0 \\
\vctr{\nabla} \cdot \vctr{T}_R+\partial_t \vctr{J}_R&=\rho(1+\frac{2}{c^2}\Phi)\vctr{G}+\frac{1}{c^2}\tr \vctr{T}\vctr{\nabla} \Phi.
\end{align}
\end{subequations}
In fact it will be useful to establish a slightly more general result:
\begin{theorem}\label{general-q}
The system of equations
\begin{subequations}\label{general-q-equations}
\begin{align}
\vctr{\nabla} \cdot \vctr{G}+\frac{n}{c^2}\vctr{G}\cdot \vctr{\nabla} \Phi&=-4\pi G_N \left(\rho(1-\frac{q}{c^2}\Phi)+\frac{1}{c^2}\tr\vctr{T} \right)\label{general-q-eq-1} \\
\vctr{\nabla} \times \vctr{G}-\frac{2}{c^2}\vctr{G}\times \vctr{\nabla} \Phi &=0\label{general-q-eq-2}  \\
\vctr{\nabla} \cdot \vctr{T}_R+\partial_t\vctr{J}_R&=\rho(1+\frac{m}{c^2}\Phi)\vctr{G}+\frac{1}{c^2}\tr \vctr{T}\vctr{\nabla} \Phi\label{general-q-eq-3} 
\end{align}
\end{subequations}
is consistent with $\partial_t\vctr{J}_R=0$ iff $q=2n-m$.
\end{theorem}
\begin{proof}See Appendix~\ref{proof-of-general-q}.\end{proof}
For $n=3,m=2$, this gives $q=4$, i.e.
\be
\rho_*=\Omega^4\rho.
\ee
This completes conformal gravity; it is striking that it has no free parameters other than the velocity $c$ (and the choice as to whether the new forces are attractive or repulsive). 

We can get some insight into why $\rho_*$ has the form it has by explicitly calculating the active and passive gravitational masses as functions of $q$. See Appendix~\ref{appendix-masses} for the details; the active mass is
\be
M^A = M^{\mathrm{int}}+\frac{1}{c^2}(2q-7)W
\ee
where $M^{\mathrm{int}}$ is the integrated mass density and 
\be
W=-\frac{1}{2}\int \rho \Phi
\ee
is the Newtonian gravitational potential energy of the body. For a spherically symmetric body, the passive mass is
\be
M^P=M^{\mathrm{int}}+\frac{1}{c^2}\frac{(q-1)}{3}W.
\ee

(For more general bodies, the passive mass is anisotropic for $q\neq 4$ and must be written as a tensor.) We see that the gravitational self-potential of a body both acts as a source of gravity and makes a contribution to the passive mass (this occurs both because of the nonlinearity of the field equations and because the coupling to the stress tensor indirectly induces a coupling to the potential, since for a static body they are related through the virial theorem). In general these contributions do not match, violating momentum conservation; only when $q=4$ do we have a consistent result
\be
M^A=M^P=M^{\mathrm{int}}+\frac{1}{c^2}W.
\ee
This has an odd consequence: if a body's internal structure changes, its integrated mass density and its total energy will remain constant but its gravitational energy need not, so that a body's gravitational mass may change over time. This breaks the normal scale-separation assumptions involved in going between discrete and continuous descriptions of matter: if a ``particle'' is a collection of matter viewed on scales where its substructure can be neglected, nonetheless the particle's mass may be time-dependent. If those particles are in turn collected into a cloud that we treat as a continuum, that continuum's mass density will violate the conservation equation at $O(1/c^2)$.

But as we will now see, this is not the worst problem for our theory.

\subsection{Failure of the Strong Equivalence Principle}

As I noted in section~\ref{continuumsection}, the content of the Strong Equivalence Principle is that gravitational mass equals inertial mass. The inertial mass of an extended body in our theory differs from the integrated mass density due to two effects: the explicit second-order term $3\partial_t^2 \Omega$ in the field equations means that acceleration changes the $\vctr{G}$ field, and acceleration induces time-dependent Coriolis fields that in turn act as a source for the curl of $\vctr{G}$. Both effects induce an acceleration-dependent self-force that alters the inertial mass; the net result, for a spherical body (again see Appendix~\ref{appendix-masses} for details, and for nonspherical bodies the result is again a tensor) is
\be
M^I=M^{\mathrm{int}}+\frac{10}{3c^2}W \neq M^P.
\ee

This does not render our theory inconsistent, but it does undermine its motivation: failure of the Strong Equivalence Principle means that the forces induced by $\vctr{G},\vctr{H},\Omega$ cannot even locally be regarded as inertial. And a closely related issue occurs when we consider the symmetry structure of the theory.

\subsection{Problems with symmetry}\label{galileansymmetry}

The equation of motion for matter in our theory was constructed by asking what equation is satisfied by inertially moving matter in conformally transformed coordinates; that led us to the fields $\vctr{G}$, $\vctr{H}$, $\Omega$. And the field equations in vacuum were then constructed by finding local equations that were satisfied by any $\vctr{G}$, $\vctr{H}$, $\Omega$ that represented a conformal transformation of empty spacetime. But it does not follow from this alone that the equation of motion, and the field equations, are covariant under time-dependent conformal transformations even when $\vctr{G}$, $\vctr{H}$, $\Omega$ are not globally inertial.

In fact, the equation of motion \emph{is} covariant for any $\vctr{G}$, $\vctr{H}$, $\Omega$: that is the content of Theorem~\ref{thm1}. But the field equations are not. In fact, they are not invariant even under time-independent conformal transformations: for a stationary distribution of mass the field equation for $\vctr{\nabla}\cdot\vctr{G}$ is
\be
\Omega^3 \vctr{\nabla} \cdot (\Omega^{-3}\vctr{G}) = -4\pi G_N\Omega^4 \rho
\ee
and the left side has conformal weight 0 while the right has conformal weight 1. The choice of source field $\rho_*=\Omega^4 \rho$, required for momentum conservation, breaks conformal symmetry. (Conversely, we could have chosen $q=3$ in section~\ref{field-equations-conformal}, and preserved conformal symmetry at the cost of violating momentum conservation.)

If we consider time-dependent conformal transformations, even the vacuum equations are not covariant.

\begin{theorem}\label{inf-conf-sym}
Under an infinitesimal conformal transformation that is trivial at some reference time and has conformal flow $\vctr{u}$ where $\partial_iu_j=O(1/c^2)$ at that time, the objects
\begin{align}
E1&=\vctr{\nabla} \cdot \vctr{C}\nonumber \\
E2&=\vctr{\nabla} \cdot \vctr{H}\nonumber \\
\vctr{E3}&=\vctr{\nabla} \times \vctr{H}+4\partial_t \vctr{C}\nonumber \\
E4&=\vctr{\nabla} \cdot \vctr{G}-3 \vctr{C}\cdot\vctr{G} -3\partial_t^2\Omega\nonumber \\
\vctr{E5}&=\vctr{\nabla} \times \vctr{G}-2\vctr{C}\times\vctr{G}+\partial_t\vctr{H}\nonumber 
\end{align}
transform to 1PN like
\begin{subequations}
\begin{align}
\delta E1&=0\\
\delta E2&=0\\
\delta \vctr{E3}&=-2\vctr{u}E1-2(\vctr{u}\cdot\vctr{\nabla})\vctr{C}\\
\delta E4&=\vctr{u}\cdot\vctr{E3}\\
\delta\vctr{E5}&=-\vctr{u}E2
\end{align}
\end{subequations}
provided that $\vctr{H}$, $(\Omega-1)$, $\vctr{C}$, $\vctr{\nabla}\times\vctr{G}$ are all $O(1/c^2)$.
\end{theorem}
\begin{proof}See Appendix~\ref{proof-of-inf-conf-sym}.\end{proof}
This is tantalizingly close to covariant, but the convective term $(\vctr{u}\cdot\vctr{\nabla})\vctr{C}$ refuses to cancel. Notice that it is nonzero even when $\vctr{u}$ is constant, so that our theory is not even Galilean-covariant.
The two problems in our theory are related: covariance under time-dependent translations was what allowed us to transform away local gravitational fields, which is exactly what the failure of the Strong Equivalence Principle tells us is impossible.

What to do? We cannot drop the conformal gradient field: as we have seen, the ratio of couplings of conformal-gradient field and rotation field to matter is fixed by mass conservation, and so if we lose one, we lose the other and are back to Newton--Cartan theory, where matter sources only the translational field. Nor can we follow the electromagnetic example and seek a redefinition of boost symmetries: the equation of motion \eqref{conformalequation} is exactly Galilean-invariant, which should make us skeptical that it will also be invariant under deformations of the Galilei group, and in fact that skepticism is well founded, as we shall see. We could accept the falsity of the Relativity Principle and formulate our theory in a preferred rest frame (a `gravitational aether', if you will), but that undermines the whole motivation of a project that started off by localizing inertial spacetime structure. Or we could seek, once again, to  cure our theory's defects by localizing more of the global structure of spacetime.

\section{Temporal scale and the Relativity Principle}\label{temporalscale}

We have already localized spatial scale; why not temporal scale, too? It will turn out that this is exactly what we need to fix the problems in our theory.

\subsection{Temporal rescaling and 1PN gravity}

Under the transformation $t\rightarrow T(t)$, it is easy to prove that the equation of motion for an inertially moving particle is
\be
 \dbd{\vctr{v}}{t} =\frac{\vctr{v}}{A}\dbd{A}{t}
\ee
where $A=(\partial_t T)^{-1}$. This can be entirely absorbed into a further mass rescaling. If we also allow for an arbitrary \emph{non}-inertial force $\vctr{F}$ (expressed in the transformed coordinates), the equation of motion for a particle under a combined conformal transformation and time rescaling is 
\be\label{generalequation}
\ddt\left(\frac{m}{A\Omega^2} \vctr{v}\right)  =\frac{1}{A\Omega^2}\vctr{F}+\left(\frac{m}{A\Omega^2}\right)(\vctr{G}+\vctr{v}\times \vctr{H} - v^2 \vctr{C})
\ee
This is not the most general inertial-force equation we could write down: we could consider quite arbitrary spatial diffeomorphisms, that leave neither shape nor size invariant. As a motivational reason not to do so: transformations of this kind seem unambiguously to represent a local distortion of purely spatial structure, and not anything that could be called `inertial'. But our reasons are ulterior: what we need is a fix for the symmetry problems in the previous theory, and this requires localizing temporal scale but not angle.

The natural next step is to try to add this new field to our existing theory, but it turns out that doing so breaks the symmetries of the equations of motion:
\begin{theorem}\label{thm-A-gal}
The equation of motion \eqref{generalequation}, with $(A-1)$, $(\Omega-1)$, $\vctr{C}$, $\vctr{H}$, and $\vctr{\nabla} \times \vctr{G}$ all $O(1/c^2)$, is covariant to 1PN under Galilean boosts (and, \emph{a fortiori}, under the group \textbf{TDConf} of general time-dependent conformal transformations) only if $A$ is spatially constant and $m$ is independent of velocity to $O(1/c^2)$.
\end{theorem}
\begin{proof}See Appendix~\ref{proof-of-thm-A}.\end{proof}

So, Galilean symmetry removes any prospect of localizing $A$: it serves only to represent global time rescalings. (The connection between Galilean covariance and the velocity-independence of mass is discussed in \citep[ch.~3]{brownrelativity}.) And a spatially constant $A$ is of no use in fixing the \emph{non}-covariance of the field equations under \textbf{TDConf}.

But now another route is available to us. If we assume that $A-1$ is $O(1/c^2)$, then a non-constant $A$ would violate covariance under \textbf{TDConf} only at that order. So perhaps there is another symmetry group, a deformation  of \textbf{TDConf} that deviates from it only to $O(1/c^2)$, such that both the field equations and the equation of motion are exactly symmetric under this new symmetry (or at least: symmetric up to terms of $O(1/c^4)$).

What can we say about this conjectured group? \textbf{TDConf} is generated by time-dependent rotations, translations, scale transformations, and special conformal transformations, and I will assume that only the time-dependent part is in need of modification. But the 1PN requirement already entails that the time-dependent parts of the dilations, special conformal transformations, and rotations must be $O(1/c^2)$, which means any $O(1/c^2)$ deformation of those transformations will vanish at 1PN. That leaves time-dependent translations, for which the most general deformation looks like
\begin{subequations}\label{deformed-trans}
\begin{align}
\delta \vctr{x}(\vctr{x},t)&=\vctr{\chi}(t)\\
\delta t(\vctr{x},t)&=\xi(\vctr{x},t)/c^2
\end{align}
\end{subequations}
Such a deformation exists: its form is fixed, uniquely up to an overall scale parameter, by the requirement that the equation of motion is covariant.

\begin{theorem}\label{thm-A-def}
If $\vctr{H}$, $(\Omega-1)$, $\vctr{C}$, and $\vctr{\nabla}\times \vctr{G}$ are all $O(1/c^2)$, equation~\eqref{generalequation} is invariant to 1PN under the group \textbf{1PNSym} of symmetries, which is generated by infinitesimal transformations of the form
\begin{subequations}
\begin{align}
\delta \vctr{x}(\vctr{x},t)&=\vctr{\mathbf{\boldsymbol{\chi}}}(t)+\vctr{w}_0\times \vctr{x}+\frac{1}{c^2}(\vctr{w}_1(t)\times\vctr{x} +2\vctr{x}(\vctr{b}(t)\cdot \vctr{x})-\vctr{b}(t)x^2+s(t)\vctr{x}) \\
\delta t(\vctr{x},t)&=\frac{\theta}{c^2}(\partial_t\vctr{\chi}(t)\cdot\vctr{x} + \eta(t))+\tau
\end{align}
\end{subequations}
for any real $\theta$, if and only if
\begin{enumerate}
\item $m$ is velocity-dependent, with 
\be
m(v)=m_0\left(1+\frac{\theta v^2}{2c^2}\right)
\ee
\item
\be
A(\vctr{x},t) = 1-\frac{\theta}{c^2}\Phi(\vctr{x},t)
\ee
for some potential $\Phi$ such that $\vctr{G}=\vctr{\nabla} \Phi+O(1/c^2)$.
\end{enumerate}

 Furthermore, this is the unique family of deformations of form \eqref{deformed-trans} of \textbf{TDConf} with this property. The fields $(\vctr{G},\vctr{H},\Omega,A)$ transform with conformal weights $(0,-1,1,0)$ under the instantaneous conformal transformation defined by $\delta \vctr{x}$, and $\vctr{G},\vctr{H},A$ pick up additional time-dependent terms which, at a time $t$ when $\delta \vctr{x}(t)=0$, take the form (up to a time-independent constant term in $\delta A$ which I conventionally set to zero)
\begin{subequations}\label{eq:G-H-trans-1pn}
\begin{align}
\delta \vctr{G}&=\partial_t\vctr{u}-\vctr{u}\times\vctr{H}-\vctr{u}(2\partial_t\Omega+\partial_tA)-\frac{2\theta}{c^2}\vctr{G}\zeta\\
\delta \vctr{H}&=-2\vctr{\omega}-2(\vctr{u}\times(\vctr{C}-\theta\vctr{G}/c^2))\\
\delta A&=-\frac{\theta}{c^2}\zeta
\end{align}
\end{subequations}
where $\vctr{u}=\partial_t\delta\vctr{x}$ is the conformal flow of the transformation, $\zeta=\partial_t\vctr{u}\cdot \vctr{x}+\partial_t\eta$, and $\vctr{\omega}$ is defined by \eqref{vorticity}. Since $\delta t\neq0$ even when $\delta\vctr{x}=0$, we need to define $\delta X\equiv X'(\tilde{\vctr{x}},\tilde t)-X(\vctr{x},t).$
\end{theorem}
\begin{proof}See Appendix~\ref{proof-of-thm-A}.\end{proof}
Note that this group also includes time reparametrizations (the temporal scale transformations whose localizations gave us $A$) although they are required to be $O(1/c^2)$. If we were considering the equation of motion in isolation, we could just absorb $\theta$ into the definition of $c$ ($\theta$ and $c$ appear only in the combination $\theta/c^2$). But the value of $c$ is fixed by the coupling of matter to the gravitational fields, so there is no additional degree of freedom.

We will also need the continuum form of these equations. The presence of $1/A$ in \eqref{generalequation} tells us that inertial forces couple to the rescaled mass density $\rho_R=\rho/A\Omega^2$, but the interpretation of $\rho$ itself needs care due to the velocity dependence of mass. For a swarm of particles without local interactions (``dust''), we have
\begin{align}
\rho \times \mbox{volume of RVE}&=\sum_n m_n(v_n^2) \nonumber\\
&=\sum_n m_{0,n}(1+\theta v_n^2/2c^2)\nonumber \\
&=\mbox{(sum of rest masses)}+\frac{\theta}{c^2}\mbox{(sum of kinetic energies)}
\end{align}
In the presence of interactions between the particles, to obtain a quantity that is conserved in the absence of inertial forces we need to replace ``kinetic energies'' with ``kinetic plus internal energies'', so that in general $\rho$ is the total rest mass plus $\theta/c^2$ times the total non-gravitational energy.  (And $\vctr{J}$ and $\vctr{T}$ must be likewise reinterpreted.) With this reinterpretation, the continuum equations of motion become
\begin{subequations}\label{eq:1pn-cont-eom}
\begin{align}
\vctr{\nabla} \cdot \vctr{T}_R+\partial_t\vctr{J}_R&=\vctr{f}_R+\rho_R \vctr{G}+\vctr{J}_R\times \vctr{H}-\tr\vctr{T}_R\vctr{C}\\
\vctr{\nabla} \cdot \vctr{J}+\partial_t\rho &=(\theta/c^2)(f_0 +\vctr{J}\cdot \vctr{G})\label{eq:1pm-cont-eom-mass}
\end{align}
\end{subequations}
where $f_0$ is the rate (per unit time, per unit volume) at which non-inertial forces do work on the matter, and $X_R=X/A\Omega^2$.

$\rho$, $\vctr{J}$, and $\vctr{T}$ transform under \eqref{deformed-trans} as
\begin{subequations}\label{eq:rel-transform}
\begin{align}
\delta \rho&= \frac{2\theta}{c^2}\vctr{J}\cdot\vctr{u}\\
\delta \vctr{J}&=\rho\vctr{u}+\frac{\theta}{c^2}(\vctr{T}\cdot\vctr{u}-\vctr{J}\zeta)\label{eq:trans-1pn-J}\\
\delta T_{ij}&=J_i u_j+J_j u_i-2 \frac{\theta}{c^2}\zeta T_{ij}
\end{align}
\end{subequations}
The transformation for $\rho$ contains two $O(1/c^2)$ terms, one from velocity-depen\-dence of mass and one from the coordinate change. The $\vctr{T}\cdot\vctr{u}$ term in $\delta\vctr{J}$ is required to preserve  \eqref{eq:1pm-cont-eom-mass} and the terms in $\zeta$ for $\vctr{J}$ and $\vctr{T}$ result from the local temporal rescaling.

\subsection{The field equations and their symmetries}

As in our previous localizations of inertial forces, the next step is to seek field equations valid in the absence of matter, that describe (at least) globally inertial fields. Rather than start from scratch, we will use the fact that our new symmetry group \textbf{1PNSym} is an $O(1/c^2)$ deformation of \textbf{TDConf}. We have already found field equations suitable for $\textbf{TDConf}$, and all except the equation for $\vctr{\nabla} \cdot \vctr{G}$ are already $O(1/c^2)$ and so will remain valid to 1PN under the deformation. To pin down $\vctr{\nabla} \cdot\vctr{G}$, consider the field that describes a finite, time-dependent acceleration $\vctr{a}(t)$, evaluated for simplicity at a time when the velocity boost is zero. We can obtain this by integrating the transformation laws \eqref{eq:G-H-trans-1pn}. It is immediate from those laws that $\vctr{H}$ and $\Omega$ are trivial. If $\vctr{G}(\epsilon), A(\epsilon)$ describe the inertial fields for acceleration $\epsilon \vctr{a}$ then
\begin{subequations}
\begin{align}
\dbd{\vctr{G}}{\epsilon}&=\vctr{a}-\frac{2\theta}{c^2}\vctr{G}(\vctr{a}\cdot\vctr{x})\\
\dbd{A}{\epsilon}&=-\frac{\theta}{c^2}\vctr{a}\cdot\vctr{x}
\end{align}
\end{subequations}
whose solutions to 1PN are
\begin{subequations}
\begin{align}
\vctr{G}(\epsilon)&=\vctr{a}\epsilon-\frac{\theta\epsilon^2}{c^2}\vctr{a}(\vctr{a}\cdot\vctr{x})\\
A(\epsilon)&=1-\frac{\theta\epsilon}{c^2}\vctr{a}\cdot\vctr{x}.
\end{align}
\end{subequations}
These fields solve
\be
\vctr{\nabla} \cdot \vctr{G}-\vctr{\nabla} A\cdot\vctr{G}=0
\ee
(note that no time derivatives of $A$ are permitted or required) and that implies that the vacuum field equation for $\vctr{\nabla}\cdot\vctr{G}$ must be
\be
\vctr{\nabla} \cdot \vctr{G}-\vctr{\nabla} A\cdot\vctr{G}-3\vctr{C}\cdot\vctr{G}-3\partial_t^2\Omega=0.
\ee
Coupling to matter now proceeds essentially as in section~\ref{field-equations-conformal}, with the exception of this change to the equation for $\vctr{\nabla} \cdot \vctr{G}$: we obtain
\begin{subequations}\label{fullfieldequations-1pn}
\begin{align}
\vctr{\nabla} A+\frac{\theta}{c^2}\vctr{G}&=0\\
\vctr{\nabla}\cdot \vctr{C}&=(4\pi G_N /c^2)\rho_* \label{eq:sec7-1}\\
\vctr{\nabla}\cdot \vctr{H}&=0 \label{eq:sec7-2} \\
\vctr{\nabla}\times\vctr{H}+4 \partial_t \vctr{C}&=-4(4\pi G_N /c^2)  \vctr{J} \label{eq:sec7-3}\\
\vctr{\nabla}\cdot \vctr{G}-\vctr{\nabla} A\cdot\vctr{G}-3\vctr{C}\cdot\vctr{G}- 3\partial_t^2 \Omega&=-4\pi G_N \left(\rho_* +\frac{\tr \vctr{T}}{c^2}\right)  \label{eq:sec7-4} \\
\vctr{\nabla}\times\vctr{G}-2 \vctr{C}\times \vctr{G}+\partial_t\vctr{H}&=0 \label{eq:sec7-5}
\end{align}
\end{subequations}
where the scaling factor for $\rho_*$ remains to be determined and might contain powers of $A$ as well as $\Omega$. Let's call the resultant theory \emph{1PN gravity}.

We can now check whether the equations of 1PN gravity are after all covariant under the deformed symmetry group. We start by considering the equations in vacuum, where we can establish:
\begin{theorem}\label{inf-1pn-sym}
If $\vctr{H}$, $(\Omega-1)$, $\vctr{C}$, $\vctr{\nabla}\times\vctr{G}$ are all $O(1/c^2)$ and the source-consistency condition $\partial_i (C_j+G_j/c^2)=0$ holds, then under an infinitesimal 1PN transformation
\begin{subequations}
\begin{align}
\delta \vctr{x}&=\vctr{\chi}(t)\\
\delta t&=\frac{\theta}{c^2}(\vctr{u}\cdot\vctr{x}+\eta(t))+\tau
\end{align}
\end{subequations}
with $\vctr{u}=\partial_t\vctr{\chi}$ and $\vctr{\chi}(t_0)=0$ at some reference time $t_0$, the set of objects
\begin{align}
\vctr{F0}&=\vctr{\nabla} A+\frac{\theta}{c^2}\vctr{G}\nonumber\\
F1&=\vctr{\nabla} \cdot \vctr{C}\nonumber \\
F2&=\vctr{\nabla} \cdot \vctr{H}\nonumber \\
\vctr{F3}&=\vctr{\nabla} \times \vctr{H}+4\partial_t \vctr{C}\nonumber \\
F4&=\vctr{\nabla} \cdot \vctr{G}-\vctr{\nabla} A\cdot\vctr{G}-3 \vctr{C}\cdot\vctr{G} -3\partial_t^2\Omega\nonumber \\
\vctr{F5}&=\vctr{\nabla} \times \vctr{G}-2\vctr{C}\times\vctr{G}+\partial_t\vctr{H}\nonumber 
\end{align}
transform among themselves to 1PN if and only if $\theta=1$, and in that situation
\begin{subequations}\label{vacuumvariations-1pn}
\begin{align}
\delta \vctr{F0}&=0\\
\delta F1&=0\\
\delta F2&=0\\
\delta \vctr{F3}&=-2\vctr{u}(F_1-F_4/c^2)\\
\delta F4&=\vctr{u}\cdot\vctr{F3}-\partial_t(\vctr{u}\cdot\vctr{F0})
-(2/c^2)(\partial_t\vctr{u}\cdot\vctr{x}+\partial_t\eta)F4\\
\delta\vctr{F5}&=-\vctr{u}F2+\vctr{u}\times\partial_t\vctr{F0}
\end{align}
\end{subequations}
\end{theorem}
\begin{proof}See Appendix~\ref{proof-of-inf-1pn-sym}.\end{proof}
This result establishes covariance for the vacuum field equations, since we already know that they are covariant under time-independent conformal transformations. (The extra term $\vctr{\nabla} A\cdot \vctr{G}$ transforms trivially under conformal transformations and so does not spoil covariance.) It also serves to fix the  parameter $\theta$. To investigate covariance in the presence of matter, we first recall that covariance under time-independent conformal transformations requires that $\rho_*$ has conformal weight 0, so that the general form of $\rho_*$ is
\be
\rho_*=A^{q-3}\Omega^3\rho.
\ee
Under an infinitesimal transformation with flow $\vctr{u}$ that is trivial at the reference time, the source density transforms like
\begin{align}\label{source-transform}
\delta \rho_*&=\frac{2}{c^2}\vctr{J}\cdot\vctr{u}-(q-3)\rho_*(\partial_t\vctr{u}\cdot\vctr{x}+\partial_t\eta)
\end{align}
Comparing \eqref{source-transform} and \eqref{eq:rel-transform} with \eqref{fullfieldequations-1pn} and \eqref{vacuumvariations-1pn}, we find that the full system of equations are covariant provided that $q=5$.

In conformal gravity, we took $\rho_*=\Omega^q\rho$, and found that there was a mismatch between the dictates of momentum conservation (which required $q=4$) and of conformal invariance (which requires $q=3$). Happily there is no mismatch here: if we adopt island-universe boundary conditions, so that $A=\Omega$ and $\rho_*=\Omega^q\rho$, then we can set $n=4$ (from $-\vctr{\nabla} A\cdot\vctr{G}-3\vctr{C}\cdot\vctr{G}$) and $m=3$ (from $\rho_R=\rho/A\Omega^2$) in Theorem~\ref{general-q} to determine that  $q=2n-m=5$ is exactly the value required for momentum to be conserved.

\subsection{Masses in 1PN gravity: restoring mass conservation and the Equivalence Principle}

Calculating the active and passive masses of an extended body is formally pretty much the same in 1PN gravity as in conformal gravity once the additional term in the equation for $\vctr{\nabla} \cdot \vctr{G}$ and the additional factor of $1/A$ in $\rho_R$ are included. For general $q$, we obtain (Appendix~\ref{appendix-masses}):
\begin{subequations}
\begin{align}
M^A&=M^{\mathrm{int}}+\frac{1}{c^2}(2q-9)W \mbox{ (active mass)}\\
M^P&=M^{\mathrm{int}}+\frac{1}{c^2}\frac{q-2}{3}W \mbox{ (passive mass, spherical body)}
\end{align}
\end{subequations}
But there is an important interpretative difference. The integrated mass density $M^{\mathrm{int}}=\int\rho$ is now the integral of the rest mass plus $1/c^2$ times the kinetic plus internal energies, so that gravity is sourced by, and responds to, non-gravitational as well as gravitational energy. Setting $q=5$ not only balances active and passive masses, it also ensures that different forms of energy couple to gravity to the same degree. With $q=5$, we have
\be
M^A=M^P=\mbox{rest mass + }\frac{1}{c^2}{\mbox{(total energy)}}
\ee
so that the mass of a body is now conserved and the scale-separation assumptions violated by conformal gravity are restored.

As for the inertial mass, its calculation needs to allow both for the altered definition of $\rho_R$ and for the fact that the transformation law \eqref{eq:trans-1pn-J} means that the stress tensor contributes to the inertial mass. The result (see Appendix~\ref{appendix-masses}) is
\be
M^I=M^{\mathrm{int}}+\frac{1}{c^2}W=M^A=M^P
\ee
so that the 1PN theory also satisfies the Strong Equivalence Principle, with total energy contributing to inertial and gravitational mass to the same degree.

At this point, we have succeeded in our goal: we have constructed an effective theory of local inertia that generalizes Newtonian gravity, is self-consistent to $O(1/c^2)$, is covariant under an $O(1/c^2)$ deformation of the time-dependent conformal group, and satisfies mass conservation and the Strong Equivalence Principle.

\section{The 1PN theory}\label{1pntheory}

Let's review the theory we have constructed. As we will see, its empirical predictions are at once striking and familiar.

\subsection{Equations of 1PN gravity}

The final form of our theory (as always, to 1PN) is
\begin{description}
\item[Field equations]
\begin{subequations}\label{finalfieldequations}
\begin{align}
\vctr{\nabla} A + \frac{1}{c^2}\vctr{G}&=0\\
\vctr{\nabla}\cdot \vctr{C}&=(4\pi G_N /c^2)\rho_* \\
\vctr{\nabla}\cdot \vctr{H}&=0 \\
\vctr{\nabla}\times\vctr{H}+4 \partial_t \vctr{C}&=-4(4\pi G_N /c^2)  \vctr{J}\\
\vctr{\nabla}\cdot \vctr{G}-\vctr{\nabla} A\cdot\vctr{G}-3\vctr{C}\cdot\vctr{G}- 3\partial_t^2 \Omega&=-4\pi G_N \left(\rho_* +\frac{\tr \vctr{T}}{c^2}\right) \label{final-eq-div-g}\\
\vctr{\nabla}\times\vctr{G}-2 \vctr{C}\times\vctr{G}+\partial_t\vctr{H}&=0
\end{align}
\end{subequations}
where $\vctr{C}=\vctr{\nabla} \ln \Omega$ and $\rho_*=A^2\Omega^3 \rho$.
\item[Equations of motion for continua]
\begin{subequations}
\begin{align}
\label{finalFD}
\vctr{\nabla} \cdot\vctr{T}_R+\partial_t \vctr{J}_R&=\vctr{f}_R+\rho_R \vctr{G}+\vctr{J}_R\times\vctr{H}-(\tr\vctr{T}_R)\vctr{C}\\
\vctr{\nabla} \cdot\vctr{J}+\partial_t \rho&=(1/c^2)(f_0+\vctr{J}\cdot\vctr{G})
\end{align}
\end{subequations}
where $\vctr{f}$ is non-inertial force density, $f_0$ is rate of energy increase due to non-inertial forces, and $X_R=X/A\Omega^2$.
\item[Equation of motion for particles]
\be\label{finalEOM}
\ddt\left( \frac{m_0 \gamma(v)}{A\Omega^2}\vctr{v}\right)=\frac{\vctr{F}_{\mathrm{ext}}}{A\Omega^2}+\frac{m_0 \gamma(v)}{A\Omega^2}\left(\vctr{G}+\vctr{v}\times\vctr{H}-v^2\vctr{C}\right)
\ee
where $\gamma(v)=(1+v^2/2c^2)$ and $m_0$ is the particle's inertial mass as measured in its rest frame. 
\item[Source-consistency boundary condition]
\be\label{finalboundary}
\partial_i (C_j+G_j/c^2)=0
\ee
\item[Scale assumptions]
$\vctr{H}$, $(\Omega-1)$, $(A-1)$, $\vctr{C}$, and $\vctr{\nabla} \times \vctr{G}$ are all $O(1/c^2)$.
\item[Spacetime symmetries]
The system of equations~\eqref{finalfieldequations}--\eqref{finalboundary} is covariant under the infinitesimal transformations
\begin{subequations}\label{fullsymmetries}
\begin{align}
\delta \vctr{x}(\vctr{x},t)&=\vctr{\mathbf{\boldsymbol{\chi}}}(t)+\vctr{w}_0\times \vctr{x}\nonumber \\
&+\frac{1}{c^2}(\vctr{w}_1(t)\times\vctr{x} +2\vctr{x}(\vctr{b}(t)\cdot \vctr{x})-\vctr{b}(t)x^2+s(t)\vctr{x}) \\
\delta t(\vctr{x},t)&=\frac{1}{c^2}(\partial_t\vctr{\chi}(t)\cdot\vctr{x} + \eta(t))+\tau
\end{align}
\end{subequations}
with $\vctr{G},\vctr{H},\Omega,A,\rho,\vctr{J},\vctr{T}$ transforming under conformal transformations with conformal weights $0,-1,1,0,-3,-3,-3$ and with additional time-dependent transformations that for  $\delta\vctr{x}(\vctr{x},t)=0$ at some reference time $t$ take this form at that time:
\begin{subequations}
\begin{align}
\delta \vctr{G}&=\partial_t\vctr{u}-\vctr{u}\times\vctr{H}-\vctr{u}(2\partial_t\Omega+\partial_tA)-(2/c^2)\zeta\vctr{G}\\
\delta \vctr{H}&=-2\vctr{\omega}-2(\vctr{u}\times(\vctr{C}-\vctr{G}/c^2))\\
\delta A&=-(1/c^2)\zeta\\
\delta \rho&=(2/c^2)\vctr{u}\cdot\vctr{J}\\
\delta \vctr{J}&=\vctr{u}\rho+(1/c^2)\vctr{T}\cdot\vctr{u}-(1/c^2)\zeta\vctr{J}\\
\delta T_{ij}&=J_i u_j + J_j u_i-(2/c^2)\zeta T_{ij}
\end{align}
\end{subequations}
where $\vctr{u}=\partial_t\delta\vctr{x}$ is the conformal flow, $2\vctr{\omega}=\vctr{\nabla}\times\vctr{u}$, and $\zeta=\partial_t\vctr{u}\cdot\vctr{x}+\partial_t\eta.$

\end{description}

\subsection{Features of 1PN gravity}\label{features-of-1pn-gravity}

Let me briefly review some of the distinctive features of the theory we have constructed.
\begin{description}
\item[Gravitomagnetism] Moving mass generates and responds to a Coriolis field, in a manner formally very similar to how moving charges generate and respond to magnetic fields. At 1PN the effect will be somewhat subtle, though, because those bodies also generate and respond to a translational Newtonian field, and the ratio of Coriolis to translational force is $v^2/c^2$. (In electromagnetism, we can have nonzero current but zero net charge, because positive and negative charges cancel, but gravity is universal.)
\item[Curvature of space] Although locally the spatially dependent scale factor can always be removed via a conformal transformation, globally it cannot: the $\Omega$ field represents genuine curvature of space, replacing Euclidean space with a conformally flat Riemannian space.
\item[Energy sources gravity] The translational gravitational field of an extended body is sourced not just by its rest mass but by its total energy; at the level of the continuum equations, $\vctr{G}$ couples not just to mass density but to $\tr \vctr{T}$ and to internal and kinetic energy density.
\item[Nonlinearity] $\vctr{G}$ couples to itself and to $A$ and $\Omega$. 
\item[Modification of inverse-square law]
If the translational field of a mass $M$ has form $\vctr{G}(r)=-G_N M/r^2 \hat{\vctr{r}}+\delta\vctr{G}(r)/c^2$, the field equations give
\be
\vctr{\nabla}\cdot\delta \vctr{G}=-4\frac{(G_N M)^2}{r^4}
\ee
so that
\be
\vctr{G}(r)=-\frac{G_N M}{r^2}\hat{\vctr{r}}\left(1-4\frac{G_N M}{rc^2}\right).
\ee
\item[Perihelion precession]
A test body in a very nearly circular orbit around (say) a star of mass $M$ at radius $r$, to Newtonian order, moves under an inverse-square force. The present theory modifies this by (i) modifying the inverse square law; (ii) adding a velocity-dependent conformal-gradient force; (iii) making inertial mass depend on $A$, $\Omega$, and velocity. Well-known results from celestial mechanics tell us that deviations from exact inverse-square motion lead to perihelion precession. The fractional precession per orbit will be of order $v^2/c^2$, on dimensional-analysis grounds; an exact calculation would\footnote{For reasons of space I do not carry out the somewhat involved task of explicitly calculating the three corrections and summing them: the overall result follows from the fact that the 1PN theory is indeed general relativity to 1PN order, plus the well-known result for general relativity; see, e.g., \citep[pp.~184--191]{weinbergGR}.} give
\be
\mbox{fractional precession per orbit} = 3 \frac{v^2}{c^2}
\ee
for an almost-circular orbit.

\item[Non-Newtonian trajectories for fast-moving objects]
The 1PN approximation only requires $v^2\ll c^2$ for bodies that meaningfully source the gravitational field; there is no reason not to consider the trajectories of low-mass test bodies with velocities $v\sim c$. For such a body, the magnitude of the conformal-gradient force $-m v^2 \vctr{C}$ will equal that of the translational force $m\vctr{G}$, and so the transverse part of the acceleration of these bodies will be twice that predicted by Newtonian gravity if $c^2>0$, or zero if $c^2<0$. (The longitudinal part will also be modified by velocity-dependence of mass.)
\end{description}

\subsection{Determining the velocity parameter}

Given the known value of the anomalous\footnote{The bulk of Mercury's precession is due to the gravitational influence of the other planets (primarily Venus and Jupiter) and has a satisfactory Newtonian explanation; see \citep[pp.~184--191]{weinbergGR} for further discussion and for sources of the parameters I use here.} precession of the perihelion of Mercury, our theory allows us to calculate $c$ explicitly. 
If we indulge the counterfactual\footnote{A similar calculation was in fact carried out by Gerber~\citep{Gerber1898} in the late 19th century, though from different field equations than ours; see \citep[pp.~35--37]{Wells2012} for a modern discussion.}  of imagining this calculation to have been carried out in the late 19th century: the anomalous precession is 43 arc seconds per century; Mercury's orbital period is 88 days, or 0.24 years, so that the anomalous precession is 0.103 arc seconds per orbit, or a fractional precession of $8.0 \times 10^{-8}$;
\be
v/c = \sqrt{8.0/3 \times 10^{-8}}= 1.6 \times 10^{-4}.
\ee
The orbital velocity of Mercury is $4.8 \times 10^4 \, \mathrm{m\,s^{-1}}$, so $|c|=2.9 \times 10^8\,\mathrm{m\,s^{-1}}$; a more accurate calculation that allowed for the orbital eccentricity of Mercury would get $3.00 \times 10^8\,\mathrm{m\,s^{-1}}$. Of course this matches the known value of the electromagnetic speed of light to the accuracy of the parameters used, even though no electromagnetic theory has been used.  The sign of the perihelion shift also fixes the sign of $c^2$ as positive.

If we then use our interpretation of gravity as localization of inertia to deduce that the trajectories of light rays should be arbitrarily well approximated by those of very light particles moving at speeds very close to $c$, we can deduce that light will be deflected by the Sun, and that the magnitude of the transverse deflection is twice what would be naively predicted from Newtonian mechanics. 

\section{Bootstrapping general relativity}\label{bootstrapping-gr}

Let's recap. 
\begin{itemize}
\item We began with the Einstein-inspired insight that Newtonian gravity can be thought of as localizing the translational-acceleration part of Galilean inertial structure, via the introduction of a universal force that could be locally transformed away. 
\item We then conjectured that other aspects of that inertial structure --- namely, rotational structure, and spatial 
scale --- might likewise be localized and coupled to matter. We started by considering only rotation, and found that the resultant theory was inconsistent unless the Coriolis field that localized rotation was source-free. 
\item We fixed this problem by localizing also spatial scale. On quite weak assumptions we found that the interparticle forces of any such theory would involve a new constant $c$ with the dimensions of velocity, as well as Newton's constant $G_N$, with Newtonian gravity recovered in the $c\rightarrow \infty$ limit. Seeking an effective-field-theory description of this theory valid to $O(1/c^2)$, we determined that theory's field equations were essentially unique under weak assumptions, up to the values of $G_N$ and $c$ and the sign under which $c^2$ appears in the equations. However, the resultant theory was not covariant under time-dependent conformal symmetries (and in particular was not Galilean-covariant), and violated mass conservation and the Strong Equivalence Principle.
\item If we also localized temporal scale via a scale field $A$, we could construct a unique theory that was covariant, to $O(1/c^2)$, under a deformation of the conformal symmetry group of order $1/c^2$ --- provided that $A$ took a specific form and that the inertial mass of bodies depended in a certain way on velocity. 
\item The resultant theory is unique up to $G_N$, $|c|$ and the sign of $c^2$, and makes a number of physical predictions that go beyond Newtonian gravity, including gravitomagnetism, deviation from the inverse-square law, enhanced deflection of fast-moving bodies, and perihelion precession. The observed value of the latter for Mercury fixes $c^2>0$ and determines $|c|$.
\end{itemize}
From an effective-theory perspective, the next step would be to iterate: to seek a new version of the theory accurate to $O(1/c^4)$, and so forth. But of course there is an easier way forward. If we initially choose a frame in which the inertial fields $\vctr{G},\vctr{H},(\Omega-1),(A-1)$ vanish, the inertial frames are exactly those obtained from that frame by transformations that generate no inertial forces, and the spacetime symmetry group is the group of transformations that generate those frames. If we start with the group \textbf{TDConf}, the spacetime symmetries are just the Galilean symmetries; but if we start with the deformed group \textbf{1PNSym}, we get a deformed group of spacetime symmetries generated by time-independent rotations and spacetime translations and by infinitesimal transformations of form
\begin{subequations}
\begin{align}
\delta \vctr{x}&=\vctr{u}t\\
\delta t&=\frac{1}{c^2}\vctr{u}\cdot\vctr{x}.
\end{align}
\end{subequations}
The decision to separately localize  various distinct components of inertial structure into $\vctr{G}$, $\vctr{H}$, $\Omega$, and $A$ relied on the assumption that in the absence of gravity the structure of spacetime is Galilean. But the modification we had to make to the boost symmetry tells us that if this theory was correct, the structure of spacetime in the absence of gravity is given \emph{not} by the Galilei group but by this deformed group. Assuming that $c^2>0$,  this symmetry group is of course the \emph{Poincar\'{e} group}. 

So the spacetime geometry is encoded not by the various structures of Galilean spacetime but by the Minkowski metric: formally, by a symmetric 2-tensor $g$ of signature $(-+++)$. That unified structure does not naturally fall apart into subcomponents, and so the strategy for localizing inertia is (with hindsight) obvious: allow $g$ to vary across spacetime. As is well known (see, e.g., \citep[pp.~155--159]{carrollgr}, \citep[pp.~69--74]{waldrelativitybook}), under fairly weak assumptions and given the need to recover Newtonian gravity as an approximation and Minkowski spacetime as a matter-free solution, the unique such theory thus constructed is general relativity (GR).

In turn, given GR we can consider regimes in which (with respect to some given asymptotically flat coordinate system) material velocities are slow compared to lightspeed and curvatures are small, and in this regime expand to first order in $1/c^2$. The resultant theory is the \emph{first post-Newtonian} (1PN) \emph{approximation} to GR --- the workhorse of Solar System tests of GR, more recently applied to the celestial mechanics of the Galactic core near its central black hole (see, e.g., \citep{GRAVITY2020}).

And, as I have scarcely attempted to hide, that 1PN approximation is exactly the effective theory of local inertial forces that we have been constructing. (Compare, e.g., \citep[eqs. 5.6, 6.14]{damouretal1991} or \citep[p.~376]{poissonwill}; note that the symmetry transformations \eqref{fullsymmetries} restrict to the symmetry group from \citep[pp.~382--383]{poissonwill} given the island-subsystem boundary condition.) The top-down path from GR and the bottom-up path from generalizing Newtonian gravity coincide to $O(1/c^2)$. 

\section{Some consequences}\label{consequences}

The conjecture that rotational as well as translational inertial structure should be localized and determined by matter has far-reaching consequences. Under weak assumptions it leads to the localization of spatial and temporal scale, to the replacement of Galilean with Poincar\'{e} covariance, and ultimately all the way to special and general relativity, without any consideration of electromagnetic phenomena and with the universal speed $c$ being required to parametrize the ratio of Coriolis to Newtonian attraction and determined empirically by the observations of Mercury.

What do we learn from this longer road to GR?

Firstly, it gives (further) reason to think that the path between Newtonian and Einsteinian gravity is smoother than it might look. It is common \citep[pp.~103--104]{Worrall1989}, \citep[pp.~1176--1181]{ruetschergrealism} to claim that there is a radical conceptual mismatch between Newtonian gravity  and GR. But as I noted in section~\ref{introduction}, this misses the point of the Equivalence Principle: Einstein made a discovery about how to understand Newtonian gravity, not (only) how to replace it. Newtonian gravity is a theory of localized translational inertia; GR is a theory of localized Minkowski structure. What the present theory shows is that we can naturally interpolate between them: start with Einstein's insight about equivalence, generalize from localizing only translational inertia to localizing also rotational inertia, and we are in the foothills of general relativity (and already beyond Galilean-covariant physics).  Among other things, this serves partially to defuse philosophical concerns  that Newtonian gravity is `just plain false, and radically so'~(Stanford~\citep[p.~9]{stanfordgrasp}; see also, e.g., \citep{eggnewton}).

Secondly, it demonstrates that the geometrization of gravity is primarily a conceptual matter, not just a matter of mathematical tools. The standard use of differential geometry in discussing not just GR but Newtonian versions of curved-spacetime gravity can give the impression that an understanding of the mathematics of manifolds and connections is the essential step required to geometrize gravity (as in Stachel's lovely parable of Newstein~\citep{Stachel2007Newstein}). But I have been at pains in this paper to use no mathematics beyond vector calculus: the important steps in the argument concern not a change in mathematical formalism but a reinterpretation of inertial forces. (That said, I don't want to entirely downplay the significance of mathematical tools: vector calculus itself, of course, was unknown for two centuries after Newton.)

Thirdly, it provides (again, further) insight into the rigidity of GR: that is, into how little freedom there is to construct a theory of gravitation embodying the Equivalence Principle. A wide variety of results  \citep{Lovelock1971Einstein,Wald1986SpinTwo,Burgess2004QGEverydayLife} support the claim that the Einstein field equations provide essentially the unique theory of gravity that combines local validity of special relativity with the Strong Equivalence Principle (at least up to potential higher-order corrections). The present work can be understood as another result in this vein: it is striking how restricted the space of generalizations of Newtonian gravity is, if we hold on to the idea that gravitational forces are locally inertial.

At a slightly more technical level, the approach here gives an insight into the structure of the 1PN approximation to GR. If we adopt boundary conditions where $\vctr{\nabla} A=\vctr{C}=-\vctr{G}/c^2$ to 1PN, the field equations~\eqref{finalfieldequations} and the particle equation of motion \eqref{finalEOM} can be rewritten as
\begin{subequations}\label{EManalog}
\begin{align}
\vctr{\nabla}\cdot \vctr{H}&=0 \\
\vctr{\nabla}\times\vctr{H}-\frac{4}{c^2} \partial_t \vctr{G}&=-4(4\pi G_N /c^2)  \vctr{J}\\
\vctr{\nabla}\cdot \vctr{G}+\frac{4}{c^2}\vctr{G}\cdot\vctr{G}+3\partial_t^2 \Phi/c^2&=-4\pi G_N \left((1-5\Phi/c^2)\rho +\frac{\tr \vctr{T}}{c^2}\right) \\
\vctr{\nabla}\times\vctr{G}+\partial_t\vctr{H}&=0\\
\ddt\left(\frac{m_0 \gamma(v)}{(1-\Phi/c^2)^3}\vctr{v}\right\}&=\frac{m_0 \gamma(v)}{(1-\Phi/c^2)^3}\left(\vctr{G}+\vctr{v}\times\vctr{H}+(v^2/c^2)\vctr{\nabla} \Phi\right)
\end{align}
\end{subequations}
In this form they show a striking analogy to Maxwell's equations, and this \emph{electromagnetic analogy} (see, e.g., \citep{CostaNatario2014GEM} and references therein) has often been useful in applications, but the analogy is partial and leaves some terms mysterious; it also creates the superficial impression that the equations support wave solutions, which in fact occur only at $O(1/c^5)$ (see, e.g., \citep[ch.~11]{poissonwill}). The interpretation of the 1PN equations in terms of translational, rotational and conformal-gradient fields offers a complementary way to understand them.

And finally, but perhaps most importantly, it helps understanding. Pedagogical appeal is a personal matter that can vary widely from person to person, but  I don't think I fully understood various features of the 1PN approximation until I saw how naturally it connected with a nonrelativistic understanding of local inertial structure. Bell begins his classic discussion of how to teach \emph{special} relativity thus:
\begin{quote}
I have for long thought that if I had the opportunity to teach this
subject, I would emphasize the continuity with earlier ideas. Usually it is
the discontinuity which is stressed, the radical break with more primitive
notions of space and time. Often the result is to destroy completely the
confidence of the student in perfectly sound and useful concepts already
acquired. \citep[p.~67]{bellsr}
\end{quote}
My approach here has much in common with Bell's and I hope it might share some of the same virtues. As he goes on to say \citep[p.~77]{bellsr}, `the longer road sometimes
gives more familiarity with the country'.
\section*{Acknowledgments}

For useful discussions, I'm grateful to Eleanor Knox and to audiences at the New Directions in Foundations of Physics meeting in Slovenia in Spring 2026 and at the Center for Philosophy of Science in Pittsburgh in Fall 2026.

The original draft of this paper was worked out by hand with no AI assistance, but Claude Opus 5.5 subsequently provided useful feedback on the overall logic, caught many mathematical errors, identified the issue with momentum conservation in section~\ref{generalizinginertial}, carried out the original calculations of the active, passive, and inertial masses in conformal and 1PN gravity, and assisted with copyediting and typesetting tasks.

\appendix
\refstepcounter{section}
\section*{Appendix: proofs of results}

\subsection{Elementary results about conformal transformations}

I adopt the notation of section~\ref{physics-in-conformal-frames}, and define $\vctr{S}=\vctr{\nabla} \ln S$. These are all standard results, which I include for completeness.

\begin{lemma} \label{lemma1}
For any conformal transformation, 
\be
[(\vctr{R}S)^{-1}\partial_k (\vctr{R}S)]_{ij}=\delta_{ij} S_k + \delta_{ik} S_j - \delta_{jk}S_i
\ee
\end{lemma}
\begin{proof}
Using the fact that partial derivatives commute, we have 
\be
\pbpbp{\tilde{x}_i}{x_j}{x_k}=\partial_k (R_{ij} S)=\partial_j (R_{ik} S)
\ee
and so
\be\label{RS-sym-1}
[(\vctr{R}S)^{-1}\partial_k (\vctr{R}S)]_{ij}- [(\vctr{R}S)^{-1}\partial_j (\vctr{R}S)]_{ik}=0.
\ee
But also
\be
[(\vctr{R}S)^{-1}\partial_k (\vctr{R}S)]_{ij} = S^{-1}\partial_k S \delta_{ij} + [\vctr{R}^{-1}\partial_k \vctr{R}]_{ij}
\ee
and $\vctr{R}^{-1}\partial_k \vctr{R}$ is antisymmetric (it's an infinitesimal rotation), from which follows
\be\label{RS-sym-2}
[(\vctr{R}S)^{-1}\partial_k (\vctr{R}S)]_{ij} + [(\vctr{R}S)^{-1}\partial_k (\vctr{R}S)]_{ji} =2\delta_{ij} S_k
\ee
The lemma follows from \eqref{RS-sym-1}, \eqref{RS-sym-2}, and a little algebra.
\end{proof}
\begin{lemma} \label{lemma2}$[\vctr{R}^{-1}\partial_k \vctr{R}]_{ij}= \delta_{ik} S_j - \delta_{jk}S_i$.
\end{lemma}
This is an immediate consequence of Lemma~\ref{lemma1}.

\begin{lemma}\label{lemma2b}
$\partial_i S_j=S_iS_j -\frac{1}{2}\delta_{ij}\vctr{S}\cdot\vctr{S}.$
\end{lemma}
\begin{proof}
Start with the matrix identity
\be
\partial_l (\vctr{R}^{-1}\partial_k \vctr{R}) - k\leftrightarrow l = (\vctr{R}^{-1}\partial_k \vctr{R})(\vctr{R}^{-1}\partial_lR)- k\leftrightarrow l 
\ee
and substitute in Lemma~\ref{lemma2} to obtain, after rearrangement,
\be
\delta_{ik}Q_{lj}-\delta_{jk}Q_{li}- \delta_{il}Q_{kj}+\delta_{jl}Q_{ki} =0,
\ee
where 
\be
Q_{ij}=\partial_i S_j -S_i S_j +\frac{1}{2}\delta_{ij}\vctr{S}\cdot\vctr{S}.
\ee
Contracting $i,k$ gives $Q_{lj}+\delta_{lj}\tr Q=0$; contracting again gives $\tr Q=0$ and hence $Q_{lj}=0$. 
\end{proof}

\begin{lemma}\label{lemma3}
For any time-dependent conformal transformation that is the identity at some time $t$, the following all hold at $t$:
\begin{subequations}
\begin{align}
 \partial_j u_i &=\omega_{ij}+\sigma \delta_{ij}\label{di-uj}\\
\vctr{\nabla} \times \vctr{u}&= 2\vctr{\omega}\label{curl-u}\\
\vctr{\nabla} \cdot \vctr{u}&=3 \sigma \label{div-u}\\
(\vctr{X}\cdot \vctr{\nabla})\vctr{u}&=\vctr{\omega}\times\vctr{X}+\sigma \vctr{X}\label{a-dot-grad-u}\\
 \partial_k\partial_j u_i &=\delta_{ij}\partial_k \sigma + \delta_{ik}\partial_j\sigma - \delta_{jk}\partial_i\sigma\label{in-terms-of-sigma}\\
 \partial_k \omega _{ij}&=\delta_{ik}\partial_j \sigma - \delta_{jk}\partial_i \sigma\label{dk-omega}\\
\vctr{\nabla} \times \vctr{\omega}&=2\vctr{\nabla} \sigma\label{curl-omega}\\
(\vctr{X}\cdot\vctr{\nabla}) \vctr{\omega}&=-\vctr{X}\times\vctr{\nabla} \sigma \label{u-dot-grad-omega}\\
\vctr{\nabla} \cdot \vctr{\omega}&=0\label{div-omega}\\
\vctr{\nabla} \sigma &= \partial_t \vctr{S}\label{grad-sigma}\\
\vctr{\nabla}\times\partial_t \vctr{u}&=2\partial_t\vctr{\omega}+2 \vctr{u}\times\vctr{\nabla}\sigma\label{deriv-u-curl}\\
\vctr{\nabla} \cdot\partial_t\vctr{u}&=3 \partial_t \sigma -3 \vctr{u}\cdot\vctr{\nabla}\sigma\label{deriv-u-div}
\end{align}
\end{subequations}
\end{lemma}

\begin{proof}
Again by the commutation of partial derivatives, we have
\be
\pbpbp{\tilde{x}_i}{x_j}{t}=\partial_j (\vctr{R}S\vctr{u})_i = \partial_t [\vctr{R}S]_{ij}
\ee
and so 
\be\label{eq:lemmastep}
[(\vctr{R}S)^{-1}\partial_j(\vctr{R}S)]_{ik}u_k + \partial_j u_i =\omega_{ij}+\sigma \delta_{ij}.
\ee
The first term disappears at time $t$, yielding \eqref{di-uj}; symmetrizing and antisymmetrizing then gives \eqref{curl-u}--\eqref{div-u}, and contracting with $\vctr{X}$ gives \eqref{a-dot-grad-u}.

\eqref{in-terms-of-sigma} follows from Lemma~\ref{lemma1} by observing that $\partial_j u_i$ is the Jacobian of an infinitesimal conformal transformation (or just by applying $\partial_k$ to both sides of \eqref{di-uj} and using the symmetries of the resultant expression in the same way as we did in the proof of Lemma~\ref{lemma1}, which amounts to the same thing). \eqref{dk-omega} is an immediate consequence of \eqref{di-uj} and \eqref{in-terms-of-sigma}, and the other identities for $\vctr{\omega}$ follow directly from \eqref{dk-omega}.

For \eqref{grad-sigma} we have
\be
\partial_j \sigma = \partial_j(S^{-1}\partial_t S)=-(S^{-1}\partial_j S)(S^{-1}\partial_t S) + S^{-1}\partial_j\partial_t S
\ee
and
\be
\partial_t S_j= \partial_t(S^{-1}\partial_j S)=-(S^{-1}\partial_t S)(S^{-1}\partial_j S) + S^{-1}\partial_t\partial_j S
\ee
Since $S^{-1}\partial_j S$ vanishes at time $t$, and partial derivatives commute, this establishes \eqref{grad-sigma}.

Now, differentiate \eqref{eq:lemmastep} once more with respect to time and evaluate at $t$ to obtain
\be
\partial_t[(\vctr{R}S)^{-1}\partial_j(\vctr{R}S)]_{ik}u_k+\partial_j\partial_t u_i =\partial_t\omega_{ij}+\partial_t\sigma\delta_{ij}
\ee
and then use Lemma~\ref{lemma1} to get
\be
\partial_t \partial_j u_i + (u_i\partial_j \sigma - u_j \partial_i \sigma) + \delta_{ij}(\vctr{u}\cdot \vctr{\nabla} \sigma)=\partial_t\omega_{ij}+\partial_t\sigma\delta_{ij}.
\ee
Antisymmetrizing over $i,j$ gives \eqref{deriv-u-curl}; contracting over $i,j$ gives \eqref{deriv-u-div}.
\end{proof}

\begin{lemma}\label{lemma4}
\begin{enumerate}
\item If $\vctr{V}$ is a (pseudo)vector field with conformal weight $-3$, then $\vctr{\nabla}\cdot \vctr{V}$ is a (pseudo)scalar field with conformal weight $-3$.
\item If $\vctr{V}$ is a vector field with conformal weight $-2$, then $\vctr{\nabla} \times \vctr{V}$ is a pseudovector field with conformal weight $-3$, and vice versa if $\vctr{V}$ is a pseudovector field.
\item If $f$ is a scalar field with conformal weight $0$, then $\vctr{\nabla} f$ is a vector field with conformal weight $-2$.
\item If $f$ is a scalar field with conformal weight $-1/2$, then $\nabla^2 f$ is a scalar field with conformal weight $-5/2$.

\end{enumerate}
\end{lemma}

\begin{proof}
The general expression for a transformed derivative of a vector field is
\be
\tilde{\partial}_k \tilde{V}_i=\frac{1}{S}[\vctr{R}^{-1}]_{nk}\partial_n([\vctr{R}S^{N+1}]_{ij}V_j).
\ee
But it is a standard vector-calculus result that these properties hold for conformal transformations with spatially constant $\vctr{R}$, and trivial to verify that they also do for spatially constant $S$. So for (1)--(2) it suffices to establish $\tilde{\vctr{\nabla}}\cdot\tilde{\vctr{V}}=\vctr{\nabla} \cdot \vctr{V}$ and $\tilde{\vctr{\nabla}}\times \tilde{\vctr{V}}=\vctr{\nabla}\times\vctr{V}$ at a  point $\vctr{x}$ where $R_{ij}(\vctr{x})=\delta_{ij}$ and $S(\vctr{x})=1$.

 Working at such a point for a conformal-weight-$N$ vector, we obtain
\begin{align}
\tilde{\partial}_k\tilde{V}_i &=[(\vctr{R}S)^{-1}\partial_k (\vctr{R}S)]_{ij}V_j+N S_k V_i +\partial_kV_i\nonumber\\
&=(N+1)S_k V_i + \delta_{ik}S_j V_j - V_k S_i+\partial_kV_i.
\end{align}
Symmetrizing and antisymmetrizing gives
\begin{subequations}
\begin{align}
\tilde{\partial}_i \tilde{V}_i&=(N+3)S_j V_j + \partial_i V_i\\
\tilde{\partial}_k \tilde{V}_i - \tilde{\partial}_i \tilde{V}_k&=(N+2)(S_k V_i - S_i V_k) + (\partial_k V_i - \partial_i V_k)
\end{align}
\end{subequations}
so that the non-covariant term vanishes for divergence when $N=-3$ and for curl when $N=-2$.

For (3), if $f$ is a conformal weight 0 scalar then
\be
\pbp{\tilde{f}}{\tilde{x}_i}=\pbp{x_j}{\tilde{x}_i}\partial_j f=[(\vctr{R}S)^{-1}]_{ji}\partial_j f=S^{-1}R_{ij}\partial_j f.
\ee

As for (4), if $f$ is a conformal weight $N$ scalar then
\begin{align}
\pbpbp{\tilde f}{\tilde{x}_i}{\tilde{x}_i}&=\pbp{x_k}{\tilde{x}_i}\pbp{}{x_k}\left( \pbp{x_j}{\tilde{x}_i}\pbp{S^N f}{x_j}\right)\nonumber \\
&=[(\vctr{R}S)^{-1}]_{ki} \partial_k\left( [(\vctr{R}S)^{-1}]_{ji}S^N[N S_j f + \partial_j f]\right)\nonumber \\
&=S^{-1}([\vctr{R}^{-1}\partial_k \vctr{R}]_{kj} +\partial_j)(S^{N-1}[N S_j f + \partial_j f]) \nonumber \\
&=S^{N-2}((N+1) S_j + \partial_j)([N S_j f + \partial_j f])\nonumber \\
&=S^{N-2}(N(N+1)\vctr{S}\cdot \vctr{S}f+N\vctr{\nabla} \cdot \vctr{S} f+ (2N+1)\vctr{S}\cdot\vctr{\nabla} f + \nabla^2 f)
\end{align}
where we used Lemma~\ref{lemma2} at step 4. The non-covariant term $\vctr{S}\cdot \vctr{\nabla} f$ vanishes for $N=-1/2$, and for this value of $N$ the terms in $\vctr{\nabla} \cdot \vctr{S}$ and $\vctr{S}\cdot \vctr{S}$ cancel by Lemma~\ref{lemma2b}.
\end{proof}

\subsection{Proof of Theorem~\ref{thm1}}\label{proof-of-theorem-1}

\begin{lemma}
The equation of motion \eqref{conformalequation} is covariant under time-independent conformal transformations, with the fields $(\vctr{F},\vctr{G},\vctr{H},\Omega)$ having conformal weights $(0,0,-1,1)$.
\end{lemma}
\begin{proof}
By considering Euclidean transformations and spatially independent scale transformations, and noting that $\tilde{\vctr{C}}=S^{-1}\vctr{R}(\vctr{C}+\vctr{S})$, we can see that  $\vctr{F}, \vctr{G},\vctr{H}$ must have the listed transformation properties. We now let $\Omega$ have conformal weight $N_\Omega$ and then isolate those parts of the transformed equation that depend on the spatial derivatives of $S,\vctr{R}$. On the right-hand side this is just
\be
\widetilde{\vctr{RHS}}=-\frac{1}{S^{2N_\Omega} \Omega^2}N_\Omega v^2 (\vctr{R}S)\vctr{S}
\ee
and on the left it is
\begin{align}\label{ti-lhs}
\widetilde{\vctr{LHS}}&=\frac{1}{\Omega^2}\vctr{v}\cdot \vctr{\nabla} \left(\frac{\vctr{R}S}{S^{2N_\Omega}}\right)\vctr{v}\nonumber\\
&=\frac{1}{S^{2N_\Omega}\Omega^2} v_j \vctr{R}S([(\vctr{R}S)^{-1}\partial_j(\vctr{R}S)]\cdot \vctr{v}-(2N_\Omega)(\vctr{v}\cdot\vctr{S})\vctr{v}),
\end{align}
with $\vctr{v}\equiv \dot{\vctr{x}}$. So the equation is covariant iff
\be
v_j[(\vctr{R}S)^{-1}\partial_j(\vctr{R}S)]\cdot \vctr{v}-(2N_\Omega)(\vctr{v}\cdot\vctr{S})\vctr{v}=-N_\Omega v^2 \vctr{S}.
\ee
Using Lemma~\ref{lemma1}, we can simplify this to
\be
(\vctr{v}\cdot\vctr{S})\vctr{v}(2-2N_\Omega)=(1-N_\Omega)v^2\vctr{S}
\ee
which holds iff $N_\Omega=1$.
\end{proof}

\subsubsection*{Proof of the theorem}

\begin{proof}
First, set the non-inertial force $\vctr{F}$ to zero, since it is clear that the result will hold for general $\vctr{F}$ if it holds for $\vctr{F}=0$. Now consider a transformation that is the identity at some fixed time $t_0$; write
\begin{subequations}
\begin{align}
\Omega&\rightarrow \alpha\Omega\\
\vctr{G}&\rightarrow\vctr{G}+\Delta\vctr{G} \\
\vctr{H}&\rightarrow\vctr{H}+\Delta\vctr{H}
\end{align}
\end{subequations}
The untransformed equation of motion can be written as
\be\label{eq:lemma-7-oldeq}
\frac{\Omega^2}{m}(\partial_t + \vctr{v}\cdot \vctr{\nabla})\left(\frac{m}{\Omega^2}\vctr{v}\right)= \vctr{G}+\vctr{v}\times \vctr{H} -v^2\vctr{C}
\ee
while the transformed one at $t_0$ is
\begin{align}\label{eq:lemma-7-neweq}
\frac{(\alpha\Omega)^2}{m}(\partial_t+\vctr{v}\cdot\vctr{\nabla})\left(\frac{m}{\alpha^2\Omega^2}\vctr{R}S(\vctr{u}+\vctr{v})\right)&=\vctr{G}+\Delta\vctr{G}+(\vctr{v}+\vctr{u})\times (\vctr{H}+\Delta \vctr{H})\nonumber \\
&- (u+v)^2(\vctr{C}+\vctr{\nabla} \ln \alpha)
\end{align}
where we have used the transformed derivative result
\be\label{body-deriv-fact}
(\partial_t +\vctr{u}\cdot\vctr{\nabla})\tilde{f}= \partial_t f(t_0)
\ee
Subtracting \eqref{eq:lemma-7-oldeq} from \eqref{eq:lemma-7-neweq}, using \eqref{body-deriv-fact} and the assumption of velocity-independent mass, and collecting powers of $\vctr{v}$ on the right, gives
\begin{align}
\alpha^2(\partial_t +\vctr{v}\cdot \vctr{\nabla})\left(\frac{\vctr{R}S}{\alpha^2}\right)\vctr{v}&+(\alpha \Omega)^2(\partial_t +\vctr{v}\cdot \vctr{\nabla})\left(\frac{\vctr{R}S}{(\alpha\Omega)^2}\vctr{u}\right)
\nonumber
\\
&=\Delta\vctr{G}-u^2(\vctr{C}+\vctr{\nabla} \ln\alpha)+\vctr{u}\times (\vctr{H}+\Delta \vctr{H})\nonumber\\
&+\vctr{v}\times \Delta \vctr{H}-2\vctr{u}\cdot\vctr{v} (\vctr{C}+\vctr{\nabla} \ln \alpha)\nonumber \\
&-v^2 \vctr{\nabla} \ln \alpha.
\end{align}
Now we expand the left-hand side and require parts constant, linear, and quadratic in $\vctr{v}$ to match separately. The quadratic part is
\begin{align}
-2\vctr{v}(\vctr{v}\cdot \vctr{\nabla} )\ln \alpha &=-v^2 \vctr{\nabla} \ln \alpha
\end{align}
which requires that $\alpha$ is spatially constant at time $t_0$. The linear part is then
\be
(\partial_t (\vctr{R}S) -2\partial_t \ln \alpha)\vctr{v}-2(\vctr{v}\cdot\vctr{C})\vctr{u}+(\vctr{v}\cdot\vctr{\nabla})\vctr{u}=\vctr{v}\times \Delta \vctr{H}-2(\vctr{u}\cdot\vctr{v})\vctr{C}
\ee
which simplifies to
\be
2(\sigma - \partial_t \ln \alpha)\vctr{v}+2\vctr{\omega} \times \vctr{v}=2\vctr{v}\times(\vctr{u}\times \vctr{C}) + \vctr{v}\times\Delta\vctr{H},
\ee
using $\vctr{a}\times (\vctr{b}\times\vctr{c})=\vctr{b}(\vctr{a}\cdot \vctr{c})-\vctr{c}(\vctr{a}\cdot\vctr{b})$ and \eqref{a-dot-grad-u}. This requires $\partial_t \ln \alpha=\sigma$ and $\Delta\vctr{H}=-2(\vctr{\omega}+\vctr{u}\times\vctr{C}).$
The constant part is then
\be
\partial_t\vctr{u}+\vctr{\omega}\times\vctr{u}-\sigma\vctr{u}-2 \vctr{u}\partial_t \ln \Omega=\Delta\vctr{G}-u^2 \vctr{C}+\vctr{u}\times(\vctr{H}+\Delta\vctr{H})
\ee
so that
\begin{align}
\Delta \vctr{G}&=\partial_t \vctr{u}+\vctr{\omega}\times \vctr{u}- \sigma \vctr{u}-2\vctr{u}\partial_t\ln \Omega +u^2\vctr{C}-\vctr{u}\times(\vctr{H}+\Delta\vctr{H}). 
\end{align}
\end{proof}

\subsection{Proof of Theorem~\ref{thm2}}\label{proof-of-theorem-2}

If we apply the transformations for $\vctr{G},\vctr{H},\Omega$ to flat spacetime in Cartesian coordinates (where $\vctr{G}=\vctr{H}=0$ and $\Omega=1$), we obtain
\begin{subequations}\label{flat-spacetime-1}
\begin{align}
\Omega&=\tilde{S}\\
\vctr{H}&=-2 (\tilde{\vctr{\omega}}+ \tilde{\vctr{u}}\times \tilde{\vctr{S}})\\
\vctr{G}&=\tilde{\vctr{u}}\times  (\tilde{\vctr{\omega}}+ 2 \tilde{\vctr{u}}\times \tilde{\vctr{S}})-\tilde{\sigma}\tilde{\vctr{u}}+\widetilde{\partial_t \vctr{u}} +\tilde{u}^2\tilde{\vctr{S}}
\end{align}
\end{subequations}

These comprise the most general description for inertial force fields resulting from a time-dependent conformal transformation. From them we can compute
\begin{subequations}\label{flat-spacetime-2}
\begin{align}
\partial_t \Omega&=\widetilde{\partial_t S}-(\tilde{\vctr{u}}\cdot\tilde{\vctr{\nabla}})\tilde{S}\\
\partial_t^2 \Omega&=\widetilde{\partial_t^2  S}-2\tilde{\vctr{u}}\cdot \widetilde{\vctr{\nabla} \partial_t S}+(\tilde{\vctr{u}}\cdot\tilde{\vctr{\nabla}})^2 \tilde{S}-\widetilde{\partial_t\vctr{u}}\cdot\tilde{\vctr{\nabla}}\tilde{S}\\
\partial_t\vctr{H}&=-2(\widetilde{\partial_t \vctr{\omega}}+\widetilde{\partial_t \vctr{u}}\times\tilde{\vctr{S}}+\tilde{\vctr{u}}\times\widetilde{\partial_t\vctr{S}})-\tilde{\vctr{u}}\cdot\tilde{\vctr{\nabla}}\tilde{\vctr{H}}
\end{align}
\end{subequations}

We proceed initially by considering only conformal transformations that are trivial at some time $t$ (so that transformed and untransformed coordinates coincide at that time), and looking for field equations valid at $t$. Evaluating \eqref{flat-spacetime-1}--\eqref{flat-spacetime-2} at $t$, we get
\begin{subequations}
\begin{align}
\Omega&=1\\
\vctr{H}&=-2\vctr{\omega}\\
\vctr{G}&=\partial_t\vctr{u}-\vctr{\omega}\times\vctr{u}-\sigma\vctr{u}\\
\partial_t\Omega&=\sigma\\
\partial_t^2\Omega&=\partial_t \sigma +\sigma^2 -2\vctr{u}\cdot\vctr{\nabla} \sigma\\
\partial_t\vctr{H}&=-2 \partial_t \vctr{\omega}+2\vctr{u}\cdot\vctr{\nabla}\vctr{\omega}-2\vctr{u}\times \vctr{\nabla} \sigma\nonumber \\
&=-2\partial_t \vctr{\omega}-4 \vctr{u}\times \vctr{\nabla} \sigma
\end{align}
\end{subequations}
where the last step used Lemma~\ref{lemma3}.
 We next calculate the divergence and curl of $\vctr{G}$ and $\vctr{H}$ and obtain (again using Lemma~\ref{lemma3})
\begin{subequations}
\begin{align}
\vctr{\nabla}\cdot\vctr{G}&=\partial_t\vctr{\nabla}\cdot\vctr{u}-\vctr{u}\cdot(\vctr{\nabla}\times\vctr{\omega})+\vctr{\omega}\cdot(\vctr{\nabla}\times\vctr{u}) -(\vctr{u}\cdot\vctr{\nabla})\sigma-\sigma \vctr{\nabla}\cdot\vctr{u}\nonumber\\
&=3\partial_t\sigma - 3\sigma^2+2\omega^2 -6(\vctr{u}\cdot\vctr{\nabla})\sigma\\
\vctr{\nabla} \times \vctr{G}&=\partial_t\vctr{\nabla} \times \vctr{u}+\vctr{u}(\vctr{\nabla}\cdot\vctr{\omega})-\vctr{\omega}(\vctr{\nabla}\cdot\vctr{u})+(\vctr{\omega}\cdot\vctr{\nabla})\vctr{u}-(\vctr{u}\cdot\vctr{\nabla})\vctr{\omega}\nonumber\\
&-\vctr{\nabla}\sigma\times \vctr{u}-\sigma \vctr{\nabla}\times\vctr{u}\nonumber\\
&=2\partial_t\vctr{\omega} -4\vctr{\omega} \sigma+ 4\vctr{u}\times\vctr{\nabla} \sigma\\
\vctr{\nabla} \cdot \vctr{H}&=-2\vctr{\nabla}\cdot\vctr{\omega}=0\\
\vctr{\nabla}\times \vctr{H}&=-2\vctr{\nabla}\times\vctr{\omega}=-4\vctr{\nabla} \sigma
\end{align}
\end{subequations}
These satisfy the equations
\begin{subequations}
\begin{align}
\vctr{\nabla} \cdot \vctr{G}&=3\partial_t^2\ln\Omega -3(\partial_t \ln\Omega)^2+\frac{1}{2}H^2\\
\vctr{\nabla} \times \vctr{G}&=-\partial_t \vctr{H}+2\vctr{H}\partial_t\Omega\\
\vctr{\nabla} \cdot \vctr{H}&=0\\
\vctr{\nabla} \times \vctr{H}&=-4\partial_t \vctr{\nabla} \Omega.
\end{align}
\end{subequations}

Since $\Omega$ is spatially constant at $t$, we can insert it freely under any spatial derivative; hence the following equations also hold at $t$ for the conformal transformations that are trivial at $t$:
\begin{subequations}
\begin{align}
\Omega^3\vctr{\nabla} \cdot(\Omega^{-3} \vctr{G})&=3\partial_t^2\ln\Omega -3(\partial_t \ln\Omega)^2+\frac{1}{2}H^2\\
\Omega^2\vctr{\nabla} \times (\Omega^{-2}\vctr{G})&=-\partial_t \vctr{H}+2\vctr{H}\partial_t\ln\Omega\\
\Omega^2\vctr{\nabla} \cdot (\Omega^{-2} \vctr{H})&=0\\
\Omega \vctr{\nabla} \times (\Omega^{-1}\vctr{H})&=-4\partial_t \vctr{C}.
\end{align}
\end{subequations}
By Lemma~\ref{lemma4}, both sides of each equation transform under time-independent conformal transformations as vectors or scalars of the same weight. So these equations are covariant under those transformations. Since any conformal transformation can be decomposed into the product of a time-independent transformation and a transformation that equals the identity at a given time, these equations must hold for all fields obtained from the trivial fields by a time-dependent conformal transformation.

Finally, note that for instantaneously spatially constant $\Omega$, trivially we have
\be
\Omega^{1/2}\nabla^2 \Omega^{-1/2}=0
\ee
and this equation is also conformally invariant by Lemma~\ref{lemma4}.

\subsection{Proof of Theorem~\ref{general-q}}\label{proof-of-general-q}

If we define 
\be
U_f(\vctr{x})=G_N \int\dr{^3y} \frac{f(\vctr{y})}{|\vctr{x}-\vctr{y}|}
\ee
so that $\nabla^2 U_f=-4\pi G_N f$ (and $\Phi=U_{\rho}$), then we can readily prove the identity
\be\label{potential-id}
\int f \vctr{\nabla} U_g=-\int g \vctr{\nabla} U_f.
\ee
Now, $\vctr{G}=\vctr{\nabla} \Phi+O(1/c^2)$, so from \eqref{general-q-eq-2} $\vctr{\nabla} \times \vctr{G}=O(1/c^4)$ and to 1PN we can write $\vctr{G}$ as a gradient. Splitting it into Newtonian and $O(1/c^2)$ terms,
\be
\vctr{G}=\vctr{\nabla} \Phi + \frac{1}{c^2}\vctr{\nabla} \Psi,
\ee
\eqref{general-q-eq-1} reduces to
\be
\nabla^2 \Psi + n |\vctr{\nabla} \Phi|^2 = -4\pi G_N (-q\rho \Phi + \tr \vctr{T}).
\ee
Using the identity $\nabla^2 \Phi^2 = 2(|\vctr{\nabla} \Phi|^2 + \Phi \nabla^2 \Phi)$, and choosing boundary conditions so $\Psi$ vanishes at infinity, we  get
\be
\Psi=-\frac{n}{2}\Phi^2 +(n-q)U_{\rho \Phi}+U_{\tr T}.
\ee
Now set $\partial_t\vctr{J}=0$ and integrate \eqref{general-q-eq-3} over the extent of the mass distribution to obtain
\begin{align}
0=&\int \left(\rho (m\Phi \vctr{\nabla} \Phi + \vctr{\nabla} \Psi )+\vctr{\nabla} \Phi\tr\vctr{T} \right)\nonumber \\
=& \int \left(\rho \Phi \vctr{\nabla} \Phi (m-n) +\rho(n-q)\vctr{\nabla} U_{\rho\Phi}\right)\nonumber \\
&+\int \left(\rho\vctr{\nabla} U_{\tr T}+\tr \vctr{T}\vctr{\nabla} \Phi\right)\nonumber\\
=&(m+q-2n)\int \rho \Phi \vctr{\nabla} \Phi
\end{align}
using \eqref{potential-id} in the last step. This vanishes iff $q=2n-m$. For $n=3,m=2$ this yields $q=4$.

\subsection{Proof of Theorem~\ref{inf-conf-sym} }\label{proof-of-inf-conf-sym}

From the transformation laws \eqref{eq:conf-G-H-trans} we can calculate (discarding some $O(1/c^4)$ terms)
\begin{subequations}
\begin{align}
\delta (\vctr{\nabla}\cdot\vctr{C})&=0\\
\delta (\vctr{\nabla} \cdot \vctr{H})&=0\\
\delta (\vctr{\nabla} \times \vctr{H}+4\partial_t\vctr{C})&=- \vctr{\nabla} \times(\vctr{\nabla}\times\vctr{u}) -2\vctr{u}\vctr{\nabla}\cdot \vctr{C}\nonumber \\
&-2 (\vctr{u}\cdot\vctr{\nabla})\vctr{C}+4\vctr{\nabla}\partial_t\Omega\\
\delta (\vctr{\nabla} \cdot \vctr{G}-3\vctr{C}\cdot\vctr{G}-3\partial_t^2\Omega)&=\vctr{\nabla}\cdot\partial_t\vctr{u}+\vctr{u}\cdot (\vctr{\nabla}\times\vctr{H})\nonumber \\
&-3\partial_t\sigma +4\vctr{u}\cdot \partial_t\vctr{C}\\
\delta (\vctr{\nabla} \times \vctr{G}-2\vctr{C}\times\vctr{G}+\partial_t\vctr{H})&=\vctr{\nabla} \times\partial_t\vctr{u}-\vctr{u}\vctr{\nabla} \cdot\vctr{H}-2\partial_t\vctr{\omega}
\end{align}
\end{subequations}
Using Lemma~\ref{lemma3}, these simplify to
\begin{subequations}
\begin{align}
\delta (\vctr{\nabla}\cdot\vctr{C})&=0\\
\delta (\vctr{\nabla} \cdot \vctr{H})&=0\\
\delta (\vctr{\nabla} \times \vctr{H}+4\partial_t\vctr{C})&=- 2\vctr{u}\vctr{\nabla}\cdot \vctr{C}-2 (\vctr{u}\cdot\vctr{\nabla})\vctr{C}\\
\delta (\vctr{\nabla} \cdot \vctr{G}-3\vctr{C}\cdot\vctr{G}-3\partial_t^2\Omega)&=\vctr{u}\cdot(\vctr{\nabla} \times \vctr{H}+4\partial_t\vctr{C})\\
\delta (\vctr{\nabla} \times \vctr{G}-2\vctr{C}\times\vctr{G}+\partial_t\vctr{H})&=-\vctr{u}\vctr{\nabla} \cdot\vctr{H}
\end{align}
\end{subequations}

\subsection{Proofs of Theorems~\ref{thm-A-gal} and \ref{thm-A-def}}\label{proof-of-thm-A}

Theorem~\ref{thm-A-gal} is a special case of Theorem~\ref{thm-A-def} with $\xi=0$,  so it suffices to prove Theorem~\ref{thm-A-def}.

To prove that theorem, in light of the discussion in the main text it suffices to consider transformations of form \eqref{deformed-trans}.
We can expand $m$ as a power series in $v^2$ (it can depend on velocity only through $v^2$ by rotational symmetry).
\begin{align}
m(v^2)=m_0(1+v^2\theta/2c^2 +O(1/c^4)),
\end{align}
Now write $\delta$ for the action of the transformation \eqref{deformed-trans} on any object, and put $\delta=\delta_1+\delta_2$ where $\delta_1$ is the action of the associated non-deformed transformation $\delta x_i=\chi_i(t),\delta t=0$. Using Theorem~\ref{thm1}, abbreviating $\mathrm{d}/\mathrm{d}t=D_t$, and working to 1PN, we have
\be\label{rel-eom-main-eq}
\delta_2 \vctr{a}+D_t \delta (\ln(m/A))\vctr{v}+D_t\ln(m/A)\vctr{u}=\delta_2\vctr{G}+\vctr{v}\times\delta_2\vctr{H}-v^2\delta_2\vctr{C}
\ee
where $\vctr{a}=\mathrm{d}\vctr{v}/\mathrm{d}t$.
Writing $\tilde{\vctr{x}}=\vctr{x}+\vctr{\chi}$, $\tilde{t}=t+ \xi/c^2$, and defining $\vctr{u}=\partial_t \vctr{\chi}$, we have
\begin{subequations}
\begin{align}
\mathrm{d}\tilde{x}_i &= \mathrm{d}x_i +\partial_t\chi_i\mathrm{d}t=(v_i+u_i)\mathrm{d}t\\
\mathrm{d}\tilde{t}&=\mathrm{d}t+(1/c^2)(\partial_t \xi \mathrm{d}t+ \partial_i \xi\mathrm{d}x_i)  =(1+ D_t\xi/c^2)\mathrm{d}t
\end{align}
\end{subequations}
so
\be
\tilde{v}_i\equiv\dbd{\tilde{x}_i}{\tilde{t}}=v_i(1- D_t \xi/c^2)+u_i
\ee
We can calculate $\mathrm{d}\tilde{v}_i$ similarly:
\begin{align}
\mathrm{d}\tilde{v}_i &=\mathrm{d}t \left\{a_i (1- D_t\xi/c^2)+\partial_t u_i -v_i  D_t^2\xi/c^2\right\}
\end{align}
so
\be
\tilde{\vctr{a}}=\dbd{\tilde{\vctr{v}}}{\tilde{t}}=\vctr{a}(1-2D_t\xi/c^2)+\partial_t\vctr{u}-\vctr{v}(D_t^2\xi/c^2)
\ee
and
\be
\delta_2\vctr{a}=\frac{1}{c^2}\left\{-2D_t\xi \vctr{G}-\vctr{v} D_t^2\xi\right\}
\ee
where we have used $\vctr{a}=\vctr{G}+O(1/c^2)$. Note that
\be
D_t^2\xi =\partial_t^2\xi + 2(\vctr{v}\cdot\vctr{\nabla})\partial_t \xi+(\vctr{G}\cdot\vctr{\nabla})\xi +(\vctr{v}\cdot\vctr{\nabla})^2\xi.
\ee

$A=1+O(1/c^2)$, so write
\be
A=1+\frac{A_1}{c^2}
\ee
and expand 
\begin{subequations}
\begin{align}
D_t \ln(m/A)&=\frac{1}{c^2}D_t( \theta v^2/2-A_1)\nonumber \\&=\frac{1}{c^2}(\theta\vctr{v}\cdot\vctr{G}-\partial_t A_1 -(\vctr{v}\cdot\vctr{\nabla})A_1)\\
\vctr{v}D_t\delta\ln(m/A)&=\frac{1}{c^2}\vctr{v}(\theta D_t(\vctr{u}\cdot\vctr{v})-D_t\delta A_1)\nonumber \\
&=\frac{1}{c^2}\vctr{v}( \theta(\vctr{u}\cdot\vctr{G}+\partial_t\vctr{u}\cdot\vctr{v})- \partial_t \delta A_1- (\vctr{v}\cdot\vctr{\nabla})\delta A_1)
\end{align}
\end{subequations}
again using $\vctr{a}=\vctr{G}+O(1/c^2)$. 

We now proceed to match powers of $v$ in \eqref{rel-eom-main-eq}. The only cubic term is the $\vctr{v}(\vctr{v}\cdot\vctr{\nabla})^2\xi$ piece in $D_t^2\xi$, so that term must vanish: $\xi$ must be at most linear in $\vctr{x}$. The quadratic part is
\be
\vctr{v}\cdot(-2\vctr{\nabla}\partial_t\xi - \vctr{\nabla} \delta A_1+ \theta\partial_t\vctr{u})\vctr{v}=-v^2\delta_2\vctr{C}
\ee
which requires $\vctr{\nabla}\delta A_1=\theta \partial_t\vctr{u}-2 \partial_t\vctr{\nabla}\xi$ and $\delta_2\vctr{C}=0$. The linear part is
\begin{align}\label{rel-linear-term}
c^2 \vctr{v}\times\delta_2\vctr{H}&=-2\vctr{G}(\vctr{v}\cdot\vctr{\nabla}\xi)  +\theta\vctr{u}(\vctr{G}\cdot \vctr{v})-\vctr{u}(\vctr{v}\cdot\vctr{\nabla} A_1)\nonumber\\
&+\vctr{v}\left\{ \theta\vctr{u}\cdot\vctr{G}-\partial_t\delta A_1-\vctr{G}\cdot\vctr{\nabla} \xi-\partial_t^2\xi\right\}
\end{align}
Choose $\vctr{v}$ to be orthogonal to $\vctr{G}$ and $\vctr{u}$ and take its inner product with \eqref{rel-linear-term} to get
\be\label{rel-v-term}
0= \theta\vctr{u}\cdot\vctr{G}-\partial_t\delta A_1-\vctr{G}\cdot\vctr{\nabla} \xi-\partial_t^2\xi
\ee
and hence
\be\label{rel-curl-term}
c^2 \vctr{v}\times\delta_2\vctr{H}=-2\vctr{G}(\vctr{v}\cdot\vctr{\nabla}\xi)  +\vctr{u}((\theta\vctr{G}-\vctr{\nabla} A_1)\cdot \vctr{v}).
\ee
\eqref{rel-curl-term} is solved if and only if
\be
\vctr{\nabla} \xi=\kappa(t)\vctr{u},\,\,\,\,\vctr{\nabla} A_1=(\theta-2\kappa(t)) \vctr{G},\,\,\,\delta_2\vctr{H}=2\kappa(t) \vctr{u}\times\vctr{G}/c^2
\ee
for arbitrary spatially independent $\kappa$.
\eqref{rel-v-term} then gives 
\be
\partial_t\delta A_1=(\theta-\kappa(t))(\vctr{u}\cdot\vctr{G})-\partial_t^2\xi
\ee
which forces $\kappa(t)=\theta$ since we know that $\vctr{\nabla} \delta A_1$ is independent of $\vctr{G}$. We can then solve to get
\begin{subequations}
\begin{align}
\xi(\vctr{x},t)&=\theta( \vctr{u}\cdot\vctr{x}+\eta(t))\\
A_1(\vctr{x},t)&=- \theta\Phi(\vctr{x},t)\\
\delta A_1(\vctr{x},t)&=-\theta(\partial_t\vctr{u}\cdot\vctr{x}+\partial_t\eta(t))+\mathrm{constant}.
\end{align}
\end{subequations}
Finally, the constant part is
\be
c^2\delta_2 \vctr{G}=-\vctr{u}\partial_t A_1 -2\theta\vctr{G}(\partial_t\vctr{u}\cdot\vctr{x}+\partial_t\eta).
\ee
The remainder of the proof now follows from Theorem~\ref{thm1}.

\subsection{Proof of Theorem~\ref{inf-1pn-sym} }\label{proof-of-inf-1pn-sym}

Write $\delta_2$ for the difference between the effect of the 1PN transformation and the corresponding conformal transformation: that is, $\delta_2$ comprises the terms in $A$ and $\theta/c^2$ in the transformation rules \eqref{eq:G-H-trans-1pn} and the transformations 
\begin{subequations}
\begin{align}
\vctr{\nabla}&\rightarrow\vctr{\nabla} -(\theta/c^2)\vctr{u}\partial_t \\
\partial_t&\rightarrow \partial_t-\vctr{u}\cdot \vctr{\nabla}-\zeta(\theta/c^2)\partial_t
\end{align}
\end{subequations}
 with $\zeta=\vctr{x}\cdot\partial_t\vctr{u}+\partial_t\eta$. ($\delta_2\vctr{\nabla}$ and $\delta_2\partial_t$ are trivial to 1PN except when acting on $\vctr{G}$, so in fact only $\delta_2\vctr{\nabla}$ matters since $\partial_t\vctr{G}$ does not appear in the field equations.) We can then calculate
\begin{subequations}
\begin{align}
\delta_2 (\vctr{\nabla}\cdot\vctr{C})&=0\\
\delta_2 (\vctr{\nabla} \cdot \vctr{H})&=(\theta/c^2)\vctr{u}\cdot(\vctr{\nabla}\times \vctr{G})=0 \mbox{ (to 1PN)}\\
\delta_2 (\vctr{\nabla} \times \vctr{H}+4\partial_t\vctr{C})&=(\theta/c^2)(2\vctr{u}\vctr{\nabla}\cdot\vctr{G}-2(\vctr{u}\cdot\vctr{\nabla})\vctr{G})\\
\delta_2 (\vctr{\nabla} \cdot \vctr{G}-3\vctr{C}\cdot\vctr{G}-3\partial_t^2\Omega)&=-\vctr{u}\cdot\vctr{\nabla}\partial_t A\nonumber\\
&-(\theta/c^2)(\vctr{u}\cdot\partial_t\vctr{G}+2\zeta \vctr{\nabla}\cdot\vctr{G}+2\vctr{G}\cdot\partial_t\vctr{u})\\
\delta_2(\vctr{\nabla} \times \vctr{G}-2\vctr{C}\times\vctr{G}+\partial_t\vctr{H})&=\vctr{u}\times\partial_t\vctr{\nabla} A+ (\theta/c^2)\vctr{u}\times\partial_t\vctr{G}\\
\delta (\vctr{\nabla} A\cdot\vctr{G})&=\vctr{\nabla} A\cdot \partial_t\vctr{u}-(\theta/c^2)\vctr{G}\cdot\partial_t\vctr{u}.
\end{align}
\end{subequations}
Using Theorem~\ref{inf-conf-sym}, we can then obtain
\begin{subequations}
\begin{align}
\delta \vctr{F0}&=0\\
\delta F1&=0\\
\delta F2&=0\\
\delta \vctr{F3}&=-2\vctr{u}(F1-(\theta/c^2)F4)
-2(\vctr{u}\cdot\vctr{\nabla})(\vctr{C}+(\theta/c^2)\vctr{G})\\
\delta F4&=\vctr{u}\cdot\vctr{F3}-\partial_t(\vctr{u}\cdot\vctr{F0})
-(2\theta/c^2)\zeta F4\\
\delta\vctr{F5}&=-\vctr{u}\delta F1+\vctr{u}\times\partial_t\vctr{F0}
\end{align}
\end{subequations}
from which the result follows by source-consistency.

\subsection{Masses of a body}\label{appendix-masses}
We need two lemmas from celestial mechanics, which for convenience I reproduce here. (For the first, see, e.g., \citep[pp.~211--213]{binneytremaine}; for the second, e.g., \citep[p.~333]{poissonwill}.)
\begin{lemma}[Tensor Virial Theorem]
Given a distribution of matter interacting by Newtonian gravity and bounded within a region of finite spatial extent, its stress tensor $T_{ij}$ satisfies
\be
\langle W_{ij}\rangle = -\int \langle T_{ij}\rangle
\ee
where $\langle\cdot \rangle$ denotes a time average, $W_{ij}$ is Chandrasekhar's potential energy tensor, defined by 
\be\label{eq:chand}
W_{ij}=-\frac{G_N}{2}\int \dr{^3x}\dr{^3y}\rho(\vctr{x})\rho(\vctr{y})\frac{(x_i-y_i)(x_j-y_j)}{|\vctr{x}-\vctr{y}|^3}=\int \dr{^3x}\rho(\vctr{x})x_i\partial_j\Phi(\vctr{x}),
\ee
and $\Phi$ is the Newtonian gravitational potential of the distribution. 
\end{lemma}
\begin{proof}
Define the moment of inertia tensor
\be
I_{ij}=\int\dr{^3x}\rho(\vctr{x})x^ix_j.
\ee
Differentiate it with respect to time, use $\partial_t\rho+\vctr{\nabla} \cdot \vctr{J}=0$, and then integrate by parts:
\be
\ddt I_{ij}=\int\dr{^3x}\partial_t \rho x_i x_j =-\int \dr{^3x}(\partial_k J_k)x^ix_j=\int \dr{^3x}(x_i J_j + x_j J_i).
\ee
Differentiate again, use $\partial_t \vctr{J}+\vctr{\nabla} \cdot \vctr{T}=\rho\vctr{\nabla} \Phi$, and integrate by parts again:
\begin{align}
\dbdbd{}{t}I_{ij}&=\int \dr{^3x}(x_i \partial_t J_j + x_j \partial_t J_i)\nonumber \\
&=\int\dr{^3x}(\rho(\vctr{x})x_i\partial_j \Phi -x_i\partial_kT_{kj}+\rho(\vctr{x})x_j \partial_i \Phi- x_j \partial_k T_{ki})\nonumber \\
&=2W_{ij}+2\int T_{ij}.
\end{align}
Finally, take time-averages.
\end{proof}
Note that the trace of Chandrasekhar's tensor is the gravitational potential energy $W$, so this immediately yields the scalar virial theorem, $\langle W\rangle=-\int\langle \tr \vctr{T}\rangle$. For a collection of particles interacting only under gravity, $\tr\vctr{T}$ is twice the total kinetic energy $K$, so that $\langle W \rangle=-2\langle K \rangle$; this is the form of the virial theorem used in section~\ref{oneparticle}.
\begin{lemma}\label{superpotential}
The ``superpotential'',
\be
X(\vctr{x})=\frac{G_N}{2}\int\dr{^3y}|\vctr{x}-\vctr{y}|\rho(\vctr{y}),
\ee
satisfies 
\be
\partial_i\partial_j X(\vctr{x})=\frac{G_N}{2}\int \dr{^3y}\rho(\vctr{y})\left(\frac{\delta_{ij}}{|\vctr{x}-\vctr{y}|}-\frac{(x_i-y_i)(x_j-y_j)}{|\vctr{x}-\vctr{y}|^3}\right)
\ee
and in particular
\be
\nabla^2 X=\Phi.
\ee
Hence 
\be
\int \dr{^3x}\rho(\vctr{x})\partial_i\partial_j X(\vctr{x})=W_{ij}-W\delta_{ij}.
\ee
\end{lemma}
\begin{proof}
Direct calculation.
\end{proof}

Now we again consider the general equations~\eqref{general-q-equations}.

\subsubsection*{Active mass}

Assuming a static configuration of matter with boundary conditions that make the fields vanish at infinity, we have from \eqref{general-q-eq-1},
\be
M^A =-\frac{1}{4\pi G_N}\oint \vctr{G}\cdot \mathrm{d}\vctr{S}=\int\left\{  \rho(1-\frac{q}{c^2}\Phi)+\frac{1}{c^2}\tr \vctr{T}+\frac{n}{4 \pi G_N c^2}|\vctr{\nabla} \Phi|^2\right\},
\ee
so that the gravitational field itself acts as a source, both through $q$ and through the nonlinear term in the vacuum equations. Using $\nabla^2 \Phi^2=2|\vctr{\nabla} \Phi|^2 + 2 \Phi \nabla^2 \Phi$ and eliminating a boundary term, and applying the scalar virial theorem, we get 
\be
M^A = M^{\mathrm{int}} + \frac{1}{c^2}(2q-(2n+1)) W
\ee

\subsubsection*{Passive mass}

Suppose the center of mass is at $\vctr{x}=0$, and consider its response to a background gravitational field $\vctr{G}_B(\vctr{x})=\nabla\Phi_B$ sourced by distant matter. We will assume the field is constant over the mass up to terms of $O(1/c^2)$. There will be  an associated background conformal field $\Omega_B(\vctr{x})=1-\vctr{g}\cdot\vctr{x}/c^2$ with conformal-gradient field $\vctr{C}_B(\vctr{x})=-\vctr{g}/c^2$. If the Newtonian potential of the body is $\Phi_{\mathrm{body}}$, so that its translational and conformal-gradient fields are $\vctr{G}_{\mathrm{body}}=\vctr{\nabla} \Phi_{\mathrm{body}}+O(1/c^2)$ and $\vctr{C}_{\mathrm{body}}=-\vctr{\nabla} \Phi_{\mathrm{body}}/c^2$, then the total translational field is $\vctr{G}=\vctr{G}_{\mathrm{body}}+\vctr{g}+\vctr{G}_{\mathrm{induced}}$, where $\vctr{G}_{\mathrm{induced}}$ is induced by the nonlinear interaction of self-field and background field. From \eqref{general-q-eq-1}, we have
\be
\vctr{\nabla} \cdot \vctr{G}_{\mathrm{induced}}-n\vctr{C}_{\mathrm{body}}\cdot \vctr{g}-n\vctr{C}_B\cdot \vctr{G}_{\mathrm{body}}=-4\pi G_N \rho\frac{-q\vctr{g}\cdot\vctr{x} }{c^2}
\ee
which to 1PN reduces to
\be
\vctr{\nabla} \cdot \vctr{G}_{\mathrm{induced}}=\frac{1}{c^2}\vctr{g}\cdot\left(-2n\vctr{\nabla} \Phi_{\mathrm{body}}-q \vctr{x}\nabla^2 \Phi_{\mathrm{body}}  \right).
\ee
The other field equations are linear to 1PN, so the curl of $\vctr{G}_{\mathrm{induced}}$ vanishes and we can take $\vctr{G}_{\mathrm{induced}}=(1/c^2)\vctr{\nabla} \Phi_{\mathrm{induced}}$; using $\nabla^2(\vctr{g}\cdot \vctr{x}\Phi_{\mathrm{body}})=2\vctr{g}\cdot \vctr{\nabla} \Phi_{\mathrm{body}}+(\vctr{g}\cdot\vctr{x})\nabla^2\Phi_{\mathrm{body}}$ we get
\begin{subequations}
\begin{align}
\nabla^2 \Phi_{\mathrm{induced}}&=2(q-n)\vctr{g}\cdot \vctr{\nabla} \Phi_{\mathrm{body}}-q\nabla^2 (\vctr{g}\cdot \vctr{x}\Phi_{\mathrm{body}})\\
\Phi_{\mathrm{induced}}&=2(q-n)\vctr{g}\cdot\vctr{\nabla} X - q(\vctr{g}\cdot \vctr{x})\Phi_{\mathrm{body}}\\
\partial_i \Phi_{\mathrm{induced}}&=2(q-n)g_j\partial_i\partial_j X -  qg_i \Phi_{\mathrm{body}}-q (\vctr{g}\cdot\vctr{x})\partial_i\Phi_{\mathrm{body}}.
\end{align}
\end{subequations}
Now write
\begin{align}
\vctr{F}&=\vctr{F}_{\mathrm{self}}+\vctr{M}^P\cdot \vctr{g}+O(g^2)
\end{align}
for the total force on the body, and extract the part linear in $\vctr{g}$:
\begin{align}
c^2\vctr{M}^P\cdot\vctr{g}=&\vctr{g}M^{\mathrm{int}}c^2\nonumber \\
&+ \vctr{g}\int m \rho \Phi_B\nonumber \\
&+ \int \rho (m\vctr{g}\cdot \vctr{x})\vctr{\nabla} \Phi_B\nonumber \\
& + \vctr{g}\int \tr \vctr{T}\nonumber\\
&+\int \rho \vctr{\nabla} \Phi_{\mathrm{induced}}.
\end{align}
The second and third terms can be expressed in terms of the potential tensor \eqref{eq:chand}, as can the fourth if we use the virial theorem, and the fifth can be evaluated via Lemma~\ref{superpotential}. The result is:
\begin{align}
c^2\vctr{M}^P\cdot \vctr{g}&= M^{\mathrm{int}}c^2\vctr{g}\nonumber \\
&-2m W\vctr{g}\nonumber \\
&+ m \vctr{W}\cdot \vctr{g}\nonumber \\
& -W\vctr{g}\nonumber\\
&+2nW \vctr{g}+(q-2n)\vctr{W}\cdot\vctr{g}\nonumber \\
&= (2(n-m)-1)W \vctr{g}+(q-2n+m)\vctr{W}\cdot\vctr{g},
\end{align}
so that
\be
M^P_{ij}=M^{\mathrm{int}}\delta_{ij}+\frac{1}{c^2}((2(n-m)-1)W\delta_{ij}+(q-2n+m)W_{ij}).
\ee
For a spherical body, $W_{ij}=\delta_{ij}W/3$, so that
\be
M^P=M^{\mathrm{int}}+\frac{1}{c^2}\frac{(q+4n-5m-3)}{3}W
\ee
Throughout, we have assumed $\Omega_B(0)=1$; a different value will uniformly rescale the mass.

\subsubsection*{Inertial mass}

Assume that $\Omega_B(0)=1$ and let the body be initially at rest at time $0$ with acceleration $\vctr{a}$. The current at time $t$ is obtained by a velocity boost $\vctr{u}=\vctr{a}t$, giving 
\begin{subequations}
\begin{align}
\vctr{J}&=\rho\vctr{a}t  &\mbox{(Conformal gravity)}\\
\vctr{J}&=\rho\vctr{a}t+(1/c^2)t\vctr{T}\cdot\vctr{a} +O(a^2t^2)&\mbox{(1PN gravity)}
\end{align}
\end{subequations}

Let $\vctr{G}_{\mathrm{acc}}$ be the induced translational field due to the acceleration. From the $-3\partial_t^2 \Omega$ term in \eqref{eq:sec5-4}, we have
\begin{align} \vctr{\nabla} \cdot \vctr{G}_{\mathrm{acc}}(\vctr{x},t) &=3\partial_t^2\Omega_{\mathrm{body}}(\vctr{x},t)\nonumber\\ 
&=3\partial_t^2\Omega_{\mathrm{body}}(\vctr{x}-\vctr{a}t^2/2,0)\nonumber \\ 
\vctr{\nabla} \cdot \vctr{G}_{\mathrm{acc}}(\vctr{x},0) &=\frac{3}{c^2}\vctr{a}\cdot\vctr{\nabla} \Phi_{\mathrm{body}}(\vctr{x},0)\nonumber
\\&=-3\vctr{a}\cdot \vctr{C}_{\mathrm{body}}. \end{align}
From \eqref{eq:sec5-3} we have
\be
\vctr{\nabla} \times \partial_t\vctr{H}=-4\left\{\vctr{a}\vctr{\nabla} \cdot \vctr{C}_{\mathrm{body}}-(\vctr{a}\cdot \vctr{\nabla})\vctr{C}_{\mathrm{body}}\right\}=-4\vctr{\nabla} \times(\vctr{a}\times\vctr{C}_{\mathrm{body}})
\ee
which (since $\vctr{\nabla} \cdot \vctr{H}=0$) is solved for $\partial_t\vctr{H}=-4\vctr{a}\times \vctr{C}_{\mathrm{body}}$. Substituting into \eqref{eq:sec5-5} gives 
\be
\vctr{\nabla} \times \vctr{G}_{\mathrm{acc}}=4\vctr{a}\times \vctr{C}_{\mathrm{body}}.
\ee
This can be solved using the superpotential:
\be\vctr{G}_{\mathrm{acc}}=\frac{1}{c^2}\left(4\vctr{a}\Phi_{\mathrm{body}}-\vctr{\nabla} (\vctr{a}\cdot\vctr{\nabla} X)\right).
\ee
If the body is in an external field $\vctr{G}_B$ with $\vctr{G}_B(0)=\vctr{g}$, as in the passive mass definition above, integrating $\vctr{\nabla}\cdot\vctr{T}_R+\partial_t\vctr{J}_R$ over the extent of the body will give
\be
\int\rho_R\vctr{G}_{\mathrm{acc}}+\vctr{M}^P\cdot\vctr{g}+O(g^2)=\ddt\int\vctr{J}_R
\ee
so we have
\begin{align}
\vctr{M}^I\cdot\vctr{a} &= \ddt\int\vctr{J}_R- \int\rho_R\vctr{G}_{\mathrm{acc}}\nonumber\\
&=\ddt\int\vctr{J}_R+\frac{1}{c^2}(7W\vctr{a}+\vctr{W}\cdot\vctr{a}).
\end{align}

In conformal gravity we get 
\begin{align}
\ddt\int\vctr{J}_R&=\vctr{a}\int\rho_R\nonumber\\
&=\vctr{a}\int\rho(1+2\Phi/c^2)\nonumber \\
&=\vctr{a}(M^{\mathrm{int}}-4W/c^2)
\end{align}
so that
\be
M^I_{ij}=M^{\mathrm{int}}\delta_{ij}+\frac{1}{c^2}(3W\delta_{ij}+W_{ij})
\ee
or
\be
M^I=M^{\mathrm{int}}+\frac{10}{3c^2} W
\ee
for a spherical body. 
In 1PN gravity we get instead
\begin{align}
\ddt\int\vctr{J}_R&=\vctr{a}\int\rho_R+\frac{1}{c^2}\int\vctr{T}\cdot\vctr{a}\nonumber\\
&=\vctr{a}\int\rho(1+3\Phi/c^2)-(1/c^2)\vctr{W}\cdot\vctr{a}\nonumber \\
&=\vctr{a}(M^{\mathrm{int}}-6W/c^2)-(1/c^2)\vctr{W}\cdot\vctr{a}
\end{align}
by the virial theorem, so that
\be
M^I = M^{\mathrm{int}}+W/c^2.
\ee


\providecommand{\newblock}{}

\end{document}